\documentclass[aps,prd,amsmath,amssymb,superscriptaddress,showkeys,showpacs,twocolumn,floatfix]{revtex4-2}
\usepackage[final]{graphicx}
\usepackage{hyperref}
\usepackage{amsmath}
\usepackage{bbm}
\usepackage{amsfonts}
\usepackage{amssymb}
\usepackage{latexsym}
\usepackage{graphicx}
\usepackage[english]{babel}
\usepackage{multirow}
\usepackage{float}
\usepackage{url}
\usepackage{slashed}
\usepackage{xcolor}
\usepackage[utf8]{inputenc}
\usepackage{verbatim}
\usepackage{lipsum}
\usepackage{multirow}

\usepackage[utf8]{inputenc}

\usepackage{diagbox}

\usepackage{mathtools}
\usepackage{amsfonts}
\usepackage{mathrsfs}
\usepackage{bbm}
\usepackage{slashed}

\usepackage{graphicx}
\usepackage{color}
\usepackage{array}
\usepackage{esint}
\usepackage{placeins}
\usepackage{booktabs}
\usepackage{epstopdf}
\usepackage[caption=false]{subfig}

\usepackage{xspace}
\usepackage{siunitx}
\usepackage{hyperref}
\usepackage[nameinlink]{cleveref}
\usepackage{appendix}

\usepackage{xifthen}
\usepackage{xcolor}
\hypersetup{
	colorlinks,
	linkcolor={red!75!black},
	citecolor={blue!75!black},
	urlcolor={blue!75!black}
}
\usepackage{comment}

\newcommand{\be}{\begin{equation}}
\newcommand{\ee}{\end{equation}}
\newcommand{\ba}{\begin{array}}
\newcommand{\ea}{\end{array}}

\newcommand{\tr}{{\rm tr}}

\usepackage{ulem}
\newcommand{\beq}{\begin{eqnarray}}
\newcommand{\eeq}{\end{eqnarray}}

\newcommand{\ii}{\mathrm{i}}
\newcommand{\dd}{\mathrm{d}}

\begin{document}

\title{Magnetic monopoles produced by first-order phase transition}
\author{Yuefeng Di}
\affiliation{
Institute of Theoretical Physics, 
Chinese Academy of Sciences, Beijing 100190, China}
\affiliation{School of Physical Sciences, University of Chinese Academy of Sciences (UCAS), Beijing 100049, China}

\author{Ligong Bian\thanks{Corresponding Author.}}
\email{lgbycl@cqu.edu.cn}
\affiliation{Department of Physics and Chongqing Key Laboratory for Strongly Coupled Physics, Chongqing University, Chongqing 401331, P. R. China}
\author{Rong-Gen Cai}
\email{cairg@itp.ac.cn}
\affiliation{Zhejiang Key Lab of Extreme Universe, Ningbo University, Ningbo 315211, China}
\affiliation{
Institute of Fundamental Physics and Quantum Technology, Ningbo University, Ningbo, 315211, China}

\date{\today}

\begin{abstract}
 We present the first 3D lattice simulations of magnetic monopole production from a first-order phase transition, using the Press-Ryden-Spergel method, capturing the subsequent annihilation stage and the entry into the scaling regime in an expanding universe. We establish the quantitative relationship between the initial bubble number and the residual monopole abundance, and calculate the magnetic field and gravitational wave power spectra sourced by monopole dynamics. Crucially, we find that the magnetic field spectrum displays a steep $k^{-5}$ decline beyond the monopole size $w_M$, whereas the peak of the gravitational wave spectrum aligns with the scale of the mean monopole separation $d_M$, governed by $k^3$ and $k^{-3}$ scalings on the low- and high-wavenumber sides of the peak, respectively. However, since the resulting magnetic field strength is excluded by blazar observations, magnetic monopoles cannot be the origin of primordial cosmic magnetic fields. Finally, we determine the allowed monopole mass under the constraints of the Parker bound ($m_M\lesssim2.18\times10^{11}\,\mathrm{GeV}$) and dark matter abundance ($m_M\lesssim1.58\times10^{12}\,\mathrm{GeV}$). 

\end{abstract}
\maketitle
\section{Introduction}
Spontaneous symmetry breaking (SSB) in the early universe routinely gives rise to a variety of topologically stable solitons. In particular, when a grand unified gauge group $G$ undergoes a phase transition down to a subgroup containing an explicit $U(1)$ factor, 't Hooft-Polyakov monopoles~\cite{tHooft:1974kcl, Polyakov:1974ek} are inevitably produced. A single 't Hooft-Polyakov monopole carries a fundamental magnetic charge of $e_M = 4\pi/g$ and possesses an extremely large mass, typically scaled as $m_{M} \sim v$, where $v$ represents the symmetry-breaking scale, typically the Grand Unified Theories (GUTs) scale. According to standard cosmological estimates, the creation of these superheavy defects yields an abundance that would severely overclose the universe today, which is known as the ``monopole problem"~\cite{Preskill:1979zi}. While inflation~\cite{Guth:1981uk, Linde:1981mu} can resolve this puzzle by drastically diluting the energy density, an alternative approach is to consider that SSB of GUT occurs through a strongly first-order phase transition (FOPT) with supercooling. Substantial supercooling before the phase transition, and large effective monopole mass relative to the temperature right after the transition, are proposed to suppress the final number density of monopole~\cite{Guth:1979bh}.

A FOPT proceeds via the thermal or quantum nucleation of true vacuum bubbles within a supercooled false vacuum background~\cite{Coleman:1977py, Linde:1981zj, Hindmarsh_2021, Caldwell:2022qsj, Athron_2024}. Driven by the vacuum energy difference, these bubbles expand outward at relativistic velocities and eventually undergo violent collisions. Within this framework, topological defects can be produced by the Kibble mechanism~\cite{Kibble_1976, Kibble:1980mv}, which is realized through the geodesic rule~\cite{Hindmarsh:1993av,Copeland:1996jz,Kibble:1995aa}: different bubbles carry random vacuum phases, and when they collide, the phases interpolate along geodesics in the vacuum manifold. If the closed path traced by several colliding bubbles is non-contractible in the vacuum manifold, the central region cannot be smoothly filled with vacuum values, and a topological defect forms due to the topological obstruction. Furthermore, bubble coalescence also generates magnetic fields \cite{PhysRevLett.51.1488, Di_2021, Enqvist:1993np,Yang_2022,Vachaspati:1991nm,Baym:1995fk,Grasso:1997nx}, sphalerons \cite{Di:2024gsl, Di:2025ncl, fv7q-vsbn}, and gravitational waves (GWs)~\cite{Mazumdar_2019, Caprini_2016, Caprini_2020, Caldwell:2022qsj,Roshan:2024qnv}.

In this work, we systematically investigate the monopole production through the real-time dynamics of the FOPT and the subsequent evolution of magnetic monopoles using 3D lattice numerical simulations in an expanding Friedmann-Lemaître-Robertson-Walker (FLRW) universe for the first time. Specifically, we simulate the full evolutionary history from bubble nucleation to coalescence, which dynamically generates monopoles via the geodesic rule of the Kibble mechanism~\cite{Kibble_1976}. We further trace the subsequent pairwise annihilation of these monopole-antimonopole pairs until the system enters the scaling regime, thereby establishing a quantitative relationship between the initial bubble number density and the surviving monopole abundance. The spectra of the generated magnetic field and the gravitational wave sourced by monopole dynamics are also explored. 
By incorporating observational constraints from cosmic monopole searches, dark matter relic density, and magnetic fields, we separately evaluate scenarios where the monopoles reside in either the dark or visible sectors to derive rigorous bounds on the monopole mass and corresponding temperatures of the FOPTs. 

Our results demonstrate that the previous studies overestimate the dark matter relic density from the monopoles and anti-monopoles after considering annihilation, and the GW spectra from the monopoles' evolution are far too small to be probed by current detectors. Furthermore, our results demonstrate that the magnetic field strength generated by magnetic monopoles is exceedingly weak and falls well below the lower limits inferred from blazar observations. Consequently, magnetic monopoles are ruled out as the primary origin of primordial cosmic magnetic fields, necessitating an alternative magnetogenesis scenario. 

This work adopts the natural units where $\hbar=c=1$. To reasonably simplify the simulation, and considering that the phase transition completes extremely rapidly, we neglect the effects of the cosmic fluid and temperature variations.

\section{The Model}

We consider the Georgi-Glashow model with the $SU(2)$ Higgs Lagrangian being~\cite{Georgi:1974sy}:
\begin{align}
    \mathcal{L} = -\frac14 W^{a\mu\nu}W_{\mu\nu}^a + \frac12 D^\mu\phi^a D_\mu\phi^a - \mathcal{V}(\phi)\;. \label{L}
\end{align}
Where, $W^a_{\mu\nu} = \partial_\mu W^a_\nu-\partial_\nu W^a_\mu +g\varepsilon^{abc}W_\mu^bW^c_\nu$ is the field strength tensor of the $SU(2)$ gauge field $W^a_\mu$. The adjoint scalar field $\phi^a$ is represented as $\phi = \phi^a\sigma^a/2$, where $\sigma^a$ denotes the Pauli matrices. The covariant derivative is defined as $D_\mu\phi^a = \partial_\mu\phi^a + g \varepsilon^{abc} W_\mu^b\phi^c$, and $g=0.5$ is the coupling constant. The potential $\mathcal{V}(\phi)$ featuring a potential barrier
\begin{align}
    \mathcal{V}(\phi) &=  \lambda\phi^a\phi^a\left[\frac14\phi^a\phi^a  - \frac{(v+v_B)}{3}\sqrt{\phi^a\phi^a}+ \frac{vv_B}{2} \right] \label{V}
\end{align}
facilitates a first-order phase transition, where $v=1$ is the vacuum expectation value (VEV) and $v_B=0.35v$ is the location of the potential barrier.
We incorporate the model into an expanding cosmological background using the FLRW metric:
\begin{align}
    \dd s^2=-a^2(\tau)\left(\dd \tau^2-\delta_{ij}\dd x^i\dd x^j\right)\;,\label{FLRW}
\end{align}
where $a(\tau)$ is the scale factor and $\tau$ denotes the conformal time. 
Using the temporal gauge $W_0^a=0$, we obtain a self-consistent set of dynamical equations for the fields in the FLRW spacetime: 
\begin{align}
    (a^2{\phi^a}')'-a^2D^i D_i\phi^a&=-a^4\frac{\partial\mathcal{V}}{\partial\phi^a}\;,\label{SU23D}\\
    {W^{a}_i}'' + D^j W_{ij}^a &= -a^2g\varepsilon^{abc}\phi^b(D_i\phi^c)\;.\label{W3D}
\end{align}
The prime ($'$) indicates the derivative with respect to conformal time $\tau$. We assume the expansion is driven by a radiation-dominated fluid (equation of state $\omega = 1/3$); the scale factor evolves as $a(\tau) = [1 + \mathcal{H}_0 (\tau - \tau_0)]$, where $\mathcal{H}_0$ is the initial conformal Hubble parameter, and $a_0=a(\tau_0)=1$. The discretized equations of motion are detailed in the \textit{Supplemental Material}.

The first-order phase transition proceeds through the nucleation and subsequent coalescence of vacuum bubbles. Each bubble carries a local $SU(2)$ phase; consequently, upon collision, the non-trivial $SU(2)$ vacuum configurations lead to the formation of point-like topological defects, magnetic monopoles.
The four-bubble-vertex monopole picture is a heuristic prediction built upon the geodesic rule assumption for the cosmic strings from three bubbles' collision~\cite{Hindmarsh:1993av,Copeland:1996jz,Kibble:1995aa}. The rigorous mathematical generalization to $\pi_2$ monopoles via 2-simplex constructions applies primarily to global-symmetry models \cite{Creaco:2008ey}. For gauged Higgs theories, the geodesic rule is not guaranteed to hold. Two competing contributions can coexist: topological defect formation at multi-bubble vertices, and dynamical pair-production driven by Higgs-modulus oscillations and transient local symmetry restoration in collision regions. We disentangle these two channels in our lattice simulations via spatial correlation diagnostics, monitoring of the Higgs modulus, controlled-bubble tests, and parameter scans.

Since the system initially possesses no net magnetic charge, these monopoles are produced in pairs, strictly maintaining a total magnetic charge of zero at all times. Following the completion of the phase transition, the newly formed monopoles undergo mutual attraction and pairwise annihilation until the system enters (or approaches) a scaling regime. During this stage, the most characteristic behavior is the remarkable consistency in the time evolution of the comoving mean monopole separation:~\cite{Hindmarsh:2025vxh}
\begin{align}
    d_M = \left(\frac{V}{2N_M}\right)^{1/3}\;,
\end{align}
where $V$ is the simulation volume and $N_M$ is the number of monopole-antimonopole pairs. 
In this work, we focus on establishing the quantitative relationship between the number of nucleated bubbles $N_b$ and $N_M$. The procedure for identifying monopoles on the lattice is detailed in the \textit{Supplemental Material}. 


{After SU(2) symmetry is broken to U(1), a massless vector field $A_\mu= W_\mu^a\phi^a$ is produced, with field strength~\cite{tHooft:1974kcl}
\begin{align}
    A_{\mu\nu}=W_{\mu\nu}^a\hat\phi^a - \frac{1}{g}\varepsilon^{abc}\hat\phi^aD_\mu\hat\phi^bD_\nu\hat\phi^c\;,
\end{align}
where $\hat\phi^a=\phi^a/\sqrt{\phi^b\phi^b}$. So the magnetic field is $B_i=\varepsilon_{ijk}A_{jk}/2$.
We can compute its energy spectrum:
\begin{align}
    \frac{\dd \rho_B}{\dd \ln k} = 4\pi k^3\tilde B_i(k)\tilde B_i^*(k)\;,
\end{align}
where $\tilde B_i(k)\tilde B_i^*(k) = \langle\tilde B_i(\boldsymbol{k})\tilde B_i^*(\boldsymbol{k})\rangle_{|\boldsymbol{k}|=k}$, and $B_i(\boldsymbol{k})$ is the Fourier transform of $B_i(\boldsymbol{x})$.}

Also, we focus on the gravitational waves (GWs) generated during the phase transition and the subsequent contribution from monopoles after the phase transition is complete. The evolution of the GW field is governed by the linearized Einstein equation in the transverse-traceless (TT) gauge:
\begin{align}
    h_{ij}''+2\frac{a'}{a}h_{ij}'-\nabla^2h_{ij}=a^2\frac{2}{m_\mathrm{p}^2}\Pi_{ij}^\mathrm{TT}\;, \label{EoMhij}
\end{align}
where $h_{ij}$ denotes the TT component of the metric perturbation. The source term $\Pi_{ij}^\mathrm{TT}$ is the TT part of the anisotropic stress tensor $\Pi_{ij} = a^{-2} \{ D_i\phi^a D_j\phi^a - (E_i^a E_j^a + B_i^a B_j^a) / a^2 \}$~\cite{Tranberg:2017lrx}. Here, $m_\mathrm{p} = (8\pi G)^{-1/2} \simeq 2.44 \times 10^{18}$ GeV is the reduced Planck mass. The energy density of the GWs is given by $\rho_\mathrm{GW}(\tau) = \frac{m_\mathrm{p}^2}{32 \pi a^2} \langle h'_{ij} h'_{ij} \rangle$, where $\langle \dots \rangle$ denotes the volume average. The GW energy spectrum is then defined as:
\begin{align}
    \Omega_\mathrm{GW}(k,\tau) = \frac{1}{\rho_\mathrm{c}(\tau)}\frac{\dd \rho_\mathrm{GW}(\tau)}{\dd \ln k}\;,\label{OmegaGW}
\end{align}
where $\rho_\mathrm{c} = 3 H^2 m_\mathrm{p}^2$ is the critical energy density of the universe. 


\section{Simulation setup} 
In this section, we present the initialization, the modified Press-Ryden-Spergel (PRS) method for ``growing" the size of the monopole core, as well as parameter setup for the lattice simulation.
\subsection{Initialization}
 We numerically solve the complete set of equations of motion for the scalar field \eqref{SU23D} and gauge field \eqref{W3D} on a three-dimensional lattice with periodic boundary conditions.
 Initially, we instantaneously introduce a fixed number of $N_{b}$ vacuum bubbles onto the lattice, while the gauge fields and their conjugate momenta are both set to zero. The critical bubble profile is obtained by using \texttt{FindBounce}~\cite{Guada_2020}
\begin{align}
    \phi^a(r) &= \frac v2\left[1-\tanh\left(\frac{r-R_0}{l_\mathrm{w}}\right)\right]\eta^a\;,\label{bubbleprofile} \\
    \dot\phi^a(r) &= 0\;,
\end{align}
where $\eta^a$ is a random unit vector satisfying $\eta^a\eta^a=1$, $r$ is the radial distance from the bubble center, $R_0$ is the initial bubble radius, and $l_\mathrm{w}$ denotes the bubble wall thickness. These bubbles subsequently expand and coalesce until the entire simulation volume has transitioned to the true vacuum. 

Before the phase transition is complete, we first turn off the expansion of the universe and introduce damping using the Crank-Nicolson method~\cite{crank1996} with $\sigma = 1.5$ (the detailed implementation is provided in the \textit{Supplemental Material}). The purpose of this is to allow the scalar field configuration in the box to transition stably from bubbles to monopoles, facilitating the subsequent evolution of the monopole network. When the phase transition is completed ($\phi^a\phi^a \sim 1$), we turn on the expansion and simultaneously enter the ``core growth" phase of the monopoles.

\subsection{PRS Method}
When performing simulations in an expanding universe, the shrinking of the comoving size of monopoles must be taken into account. In general, the physical size of a monopole $w_M = 1/m_H$ is determined by the mass of the scalar field $m_H = \sqrt{2\lambda}v$ (or the massive boson $m_W = gv$) and is independent of the scale factor $a$. Meanwhile, the lattice spacing $\Delta x$ is comoving, with the physical grid spacing being $a \Delta x$. Consequently, the number of lattice sites occupied by a monopole, $w_M/(a\Delta x) = (am_H\Delta x)^{-1}\propto a^{-1}$, will decrease as $a$ increases. When $w_M/(a\Delta x) < 1$, the dynamics of the monopoles become unreliable. Regarding this issue, we employ an improved PRS method~\cite{1989ApJ...347..590P} adapted for gauge theories~\cite{Bevis_2010}, which involves the topological defects—in our case, magnetic monopoles—to a ``core growth" stage before physical evolution. This ``grow" their physical core size and enables them to occupy multiple lattice sites. Concretely, this ``core growth" stage is implemented by allowing the monopole core width $w_M=1/\sqrt{2\lambda}v$ to ``grow" with the scale factor $a$. Keeping the VEV $v$ invariant, this requirement dictates that the coupling parameters $\lambda$ and $g$ decrease as $a$ increases. We elaborate on the exact implementation below.

After the phase transition completes, the expansion begins, and the system simultaneously enters the ``core growth" stage of the monopoles. During this period, the coupling constants $\lambda$ and $g$ depend on $a$ as follows:
\begin{align}
    \lambda = \lambda_0a^{-6}\;,\quad g = g_0a^{-3}\;,
\end{align}
where $\lambda_0$ and $g_0$ are constant. Thereby, as $a$ increases, the number of lattice sites occupied by a monopole also increases, $w_M/(a\Delta x)\propto a^{2}$, thereby realizing the ``core growth" of the monopoles. The advantage of this approach is that it both fixes the VEV of the scalar field and keeps the mass ratio between the scalar field and the massive bosons $m_H/m_W$ constant, thereby keeping the monopole configuration stable. It should be emphasized that Ref.~\cite{Hindmarsh:2025vxh,Hindmarsh_Rummukainen_Weir_2017} generally uses $\lambda = \lambda_0a^{-4}$ (therefore $w_M/(a\Delta x)\propto a$), whereas in our simulations, the initial $\lambda_0$ is chosen to be larger to make the bubble radius relatively smaller, thereby allowing more bubbles to fit inside the simulation box. At the onset of expansion, the monopole size is actually less than one lattice spacing. Therefore, to ``grow" the monopoles within a shorter period of time, we have chosen a more aggressive $w_M/(a\Delta x)\propto a^2$. 

\begin{figure}
    \centering
    \includegraphics[width=0.8\linewidth]{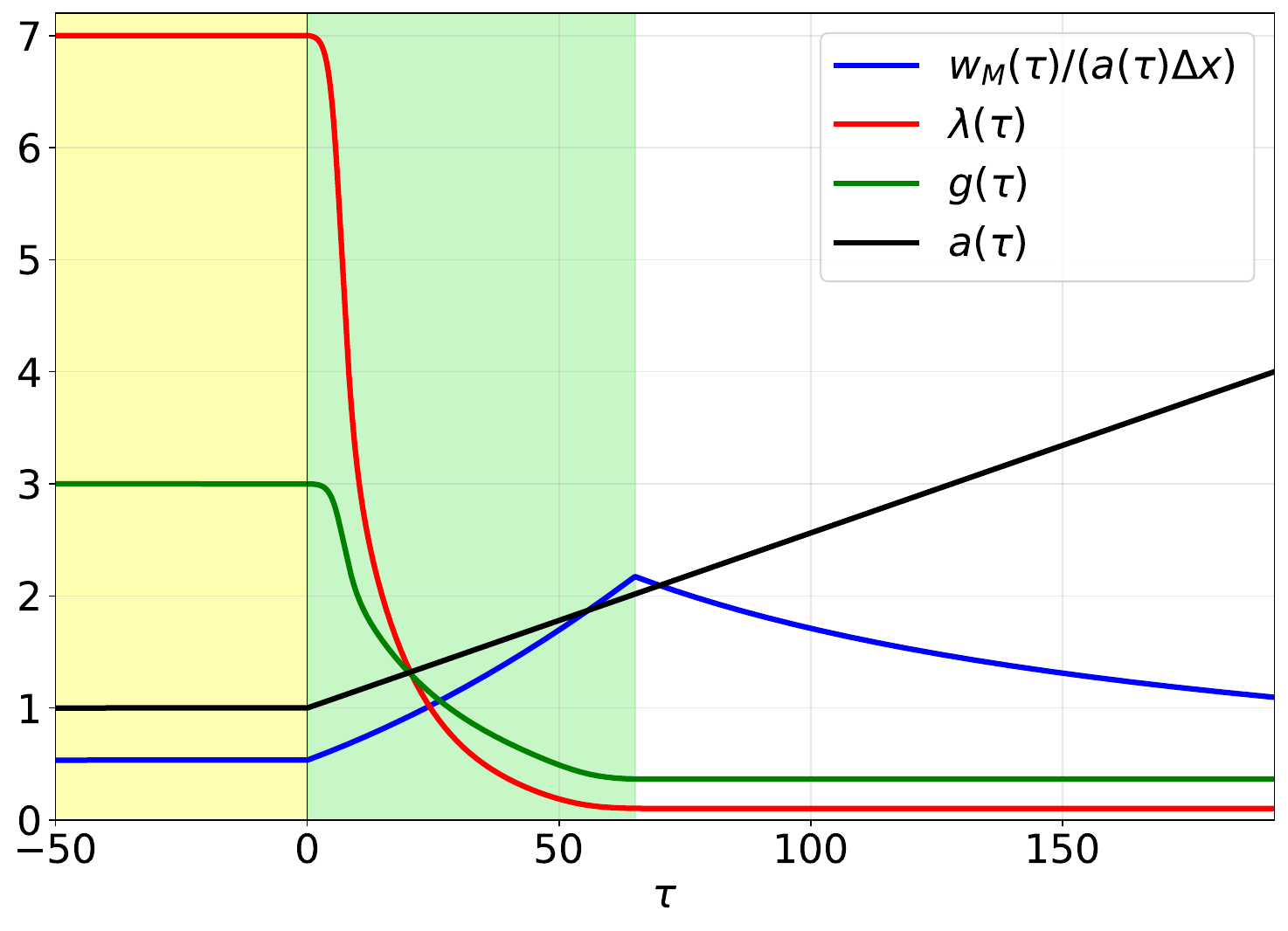}
    \caption{The time evolution of the number of lattice sites occupied by a monopole $w_M/(a\Delta x)$, the scale factor $a$, and the coupling constants $\lambda$ and $g$. We set the onset of expansion as the time $\tau = 0$. We apply smoothing functions to $\lambda$ and $g$ when entering and exiting the ``core growth" stage to make the transitions smoother. The yellow region corresponds to the phase transition stage without expansion, the green region represents the monopole ``core growth" stage, and the remaining white region denotes the physical evolution stage. }
    \label{fig:PRS}
\end{figure}
The ``core growth" stage lasts for 1300 time steps, after which $\lambda$ and $g$ are fixed at their values at that time until the end of the simulation, a period that can be referred to as the physical evolution stage. Fig. \ref{fig:PRS} shows the division of the simulation into the phase transition stage (yellow region), the ``core growth" stage (green region), and the physical evolution stage (remaining white region), as well as the time evolution of the number of lattice sites occupied by a monopole $w_M/(a\Delta x)$, the scale factor $a$, and the coupling constants $\lambda$ and $g$. At the end of the simulation, a monopole still occupies more than one lattice site. Since the number of bubbles varies, the time at which the phase transition completes also differs. For convenience in presenting the data, we define the moment when the phase transition ends and expansion begins as $\tau=\tau_0 = 0$.

When simulating an expanding universe, the number of Hubble volumes contained within the simulation box gradually decreases. We terminate the simulation when only a single Hubble volume remains inside the box.

\subsection{Parameter Setup}
For all simulations, the lattice box size is $N^3 = 512^3$, the lattice spacing is $\Delta x = 0.5$, and the time step is $\Delta \tau=0.05$. This choice of parameters provides sufficient resolution to capture all the dynamical details of both the bubbles and the monopoles. If a smaller lattice spacing $\Delta x$ were adopted, individual bubbles would occupy an excessively large number of lattice sites, preventing the simulation box from containing a large number of bubbles. Conversely, if $\Delta x$ were larger, it would be more difficult to identify the bubble walls and monopoles during the evolution, leading to a loss of dynamical details. At the beginning of the PRS stage, our simulation box encompasses 64 Hubble volumes. After evolving for 3840 time steps, only a single Hubble volume remains within the simulation box. 

For each parameter configuration listed in Table I, we perform three independent realizations and report the ensemble-averaged results.
\begin{table}[!htp]
    \centering
    \begin{tabular}{ccccccc}
        \hline\hline
        $v$ &$v_B$ & $g_0$ &$\lambda_0$ & $R_0$ & $l_\mathrm{w}$ & $N_b$ \\
        \hline
        \multirow{5}*{$1$} & \multirow{5}*{0.35} & \multirow{5}*{3} & \multirow{5}*{7} &\multirow{5}*{$10\Delta x$} &\multirow{5}*{$2\Delta x$} &$100$ \\&&&&&&$250$ \\&&&&&&$500$\\&&&&&&$750$ \\&&&&&&$1000$ \\
        \hline\hline
    \end{tabular}
    \caption{Parameters setup in the simulations. } 
    \label{tab:parameter}
\end{table}

\section{ Numerical Results}
In this section, we investigate the evolution of the monopole network following the completion of the phase transition, along with the resulting magnetic fields and gravitational waves. Detailed results regarding the phase transition dynamics itself are provided in the \textit{Supplemental Material}.

\subsection{Evolution of Monopole Anti-Monopole Network}

\begin{figure}[!htp]
    \centering
    \includegraphics[width=0.8\linewidth]{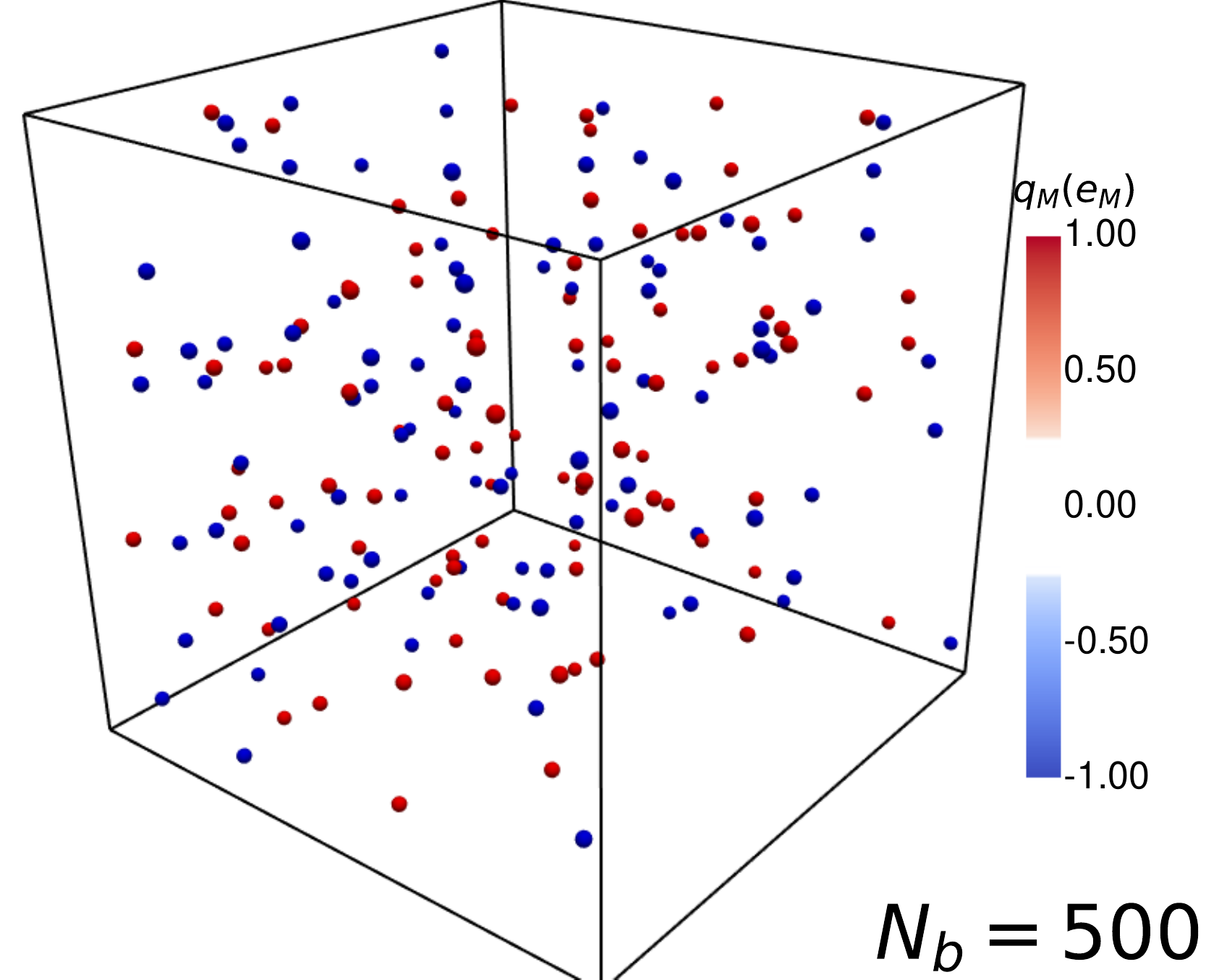}\\
    \includegraphics[width=0.8\linewidth]{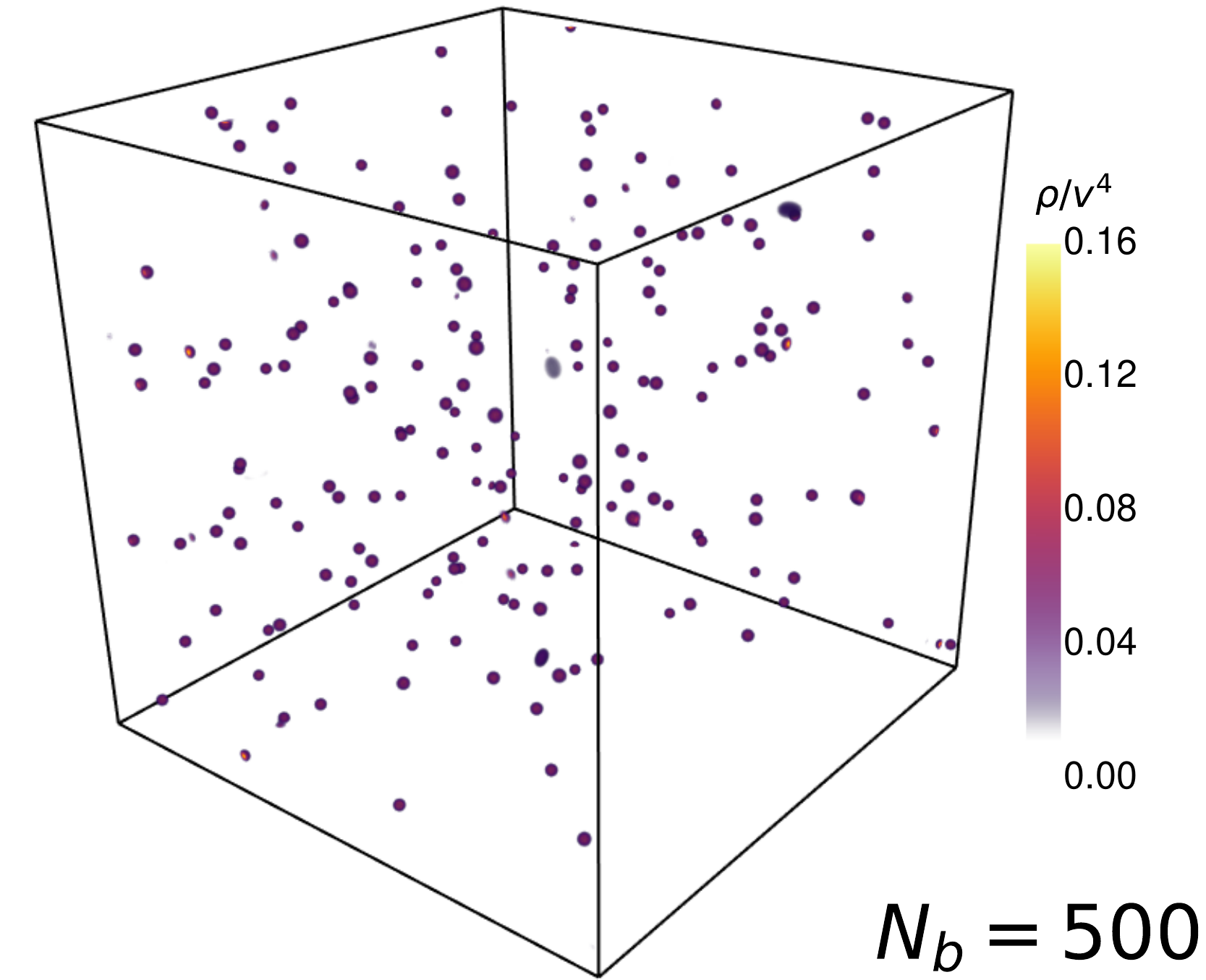}
    \caption{3D spatial distribution snapshots of the magnetic charge (Top) and the energy density configuration (Bottom) at the conclusion of the PRS stage for an initial bubble number of $N_b=500$. For clarity of visualization, the (anti)monopoles in the top panel have been magnified by a factor of 216. Each monopole carries a positive magnetic charge of $+1$ (indicated in red), while each antimonopole carries a negative charge of $-1$ (indicated in blue). 
    } 
    \label{fig:mono3D}
\end{figure}
The top panel of Fig. \ref{fig:mono3D} displays the spatial distributions of the magnetic charges at the conclusion of the simulation for an initial bubble number of $N_b=500$. For the purpose of clear visualization, the (anti)monopoles in the top panel have been magnified by a factor of 216. The simulated volume contains 92 monopoles (highlighted in red) and 92 antimonopoles (highlighted in blue) at the end of the PRS stage. These topological defects carry magnetic charges of $\pm e_M$, respectively, which strictly satisfy the conservation law of zero net magnetic charge inside the box. The bottom panel of Fig. \ref{fig:mono3D} displays the spatial distribution of the energy density at the end of the PRS stage. Compared with the top panel, it can be seen that the energy is concentrated inside the (anti)monopoles. However, there are also some regions, such as the middle right of the box, where the energy density is high but there is no magnetic charge; these are precisely the regions where monopoles and antimonopoles have annihilated. 

\begin{figure}
    \centering
    \includegraphics[width=0.8\linewidth]{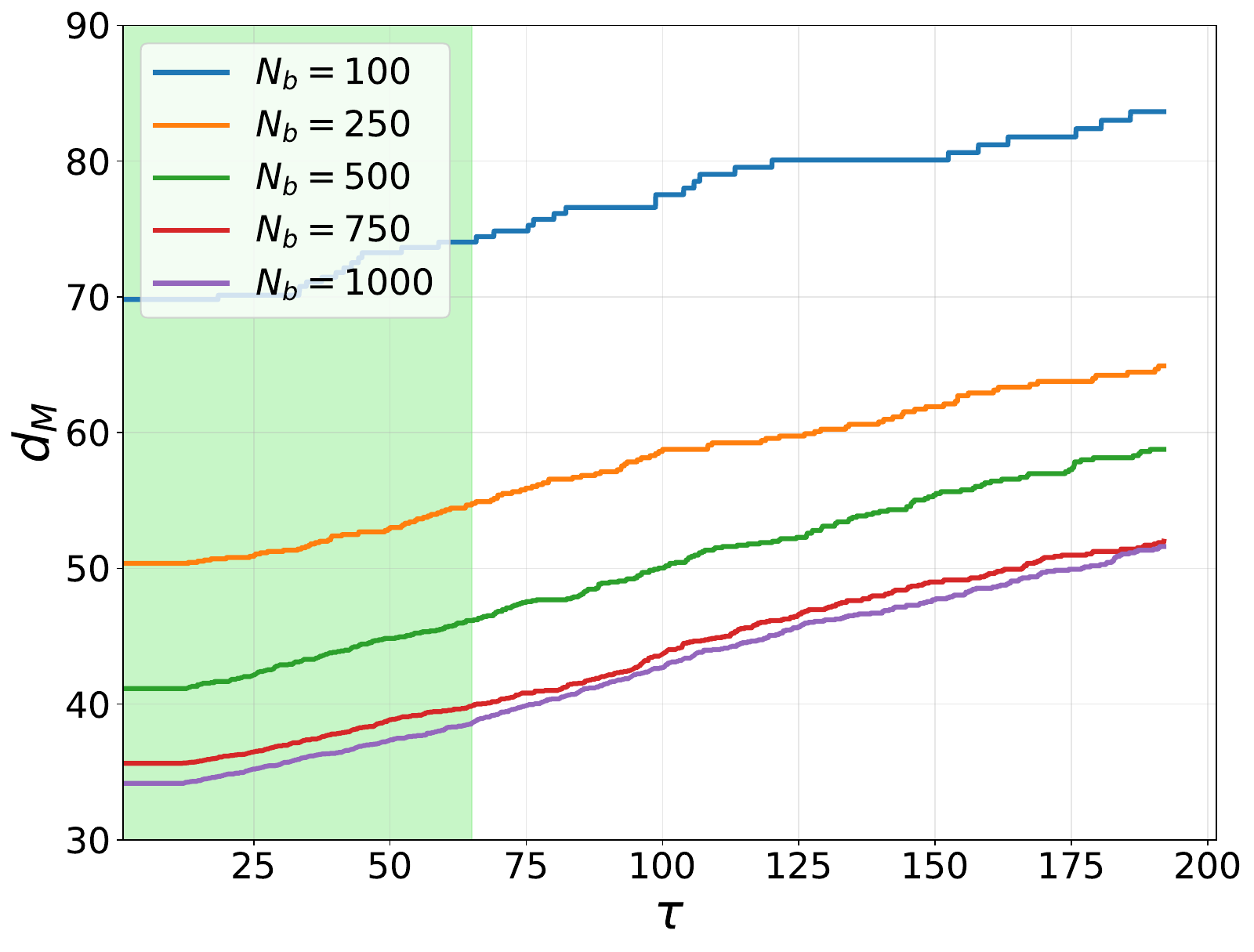}\\
    \includegraphics[width=0.8\linewidth]{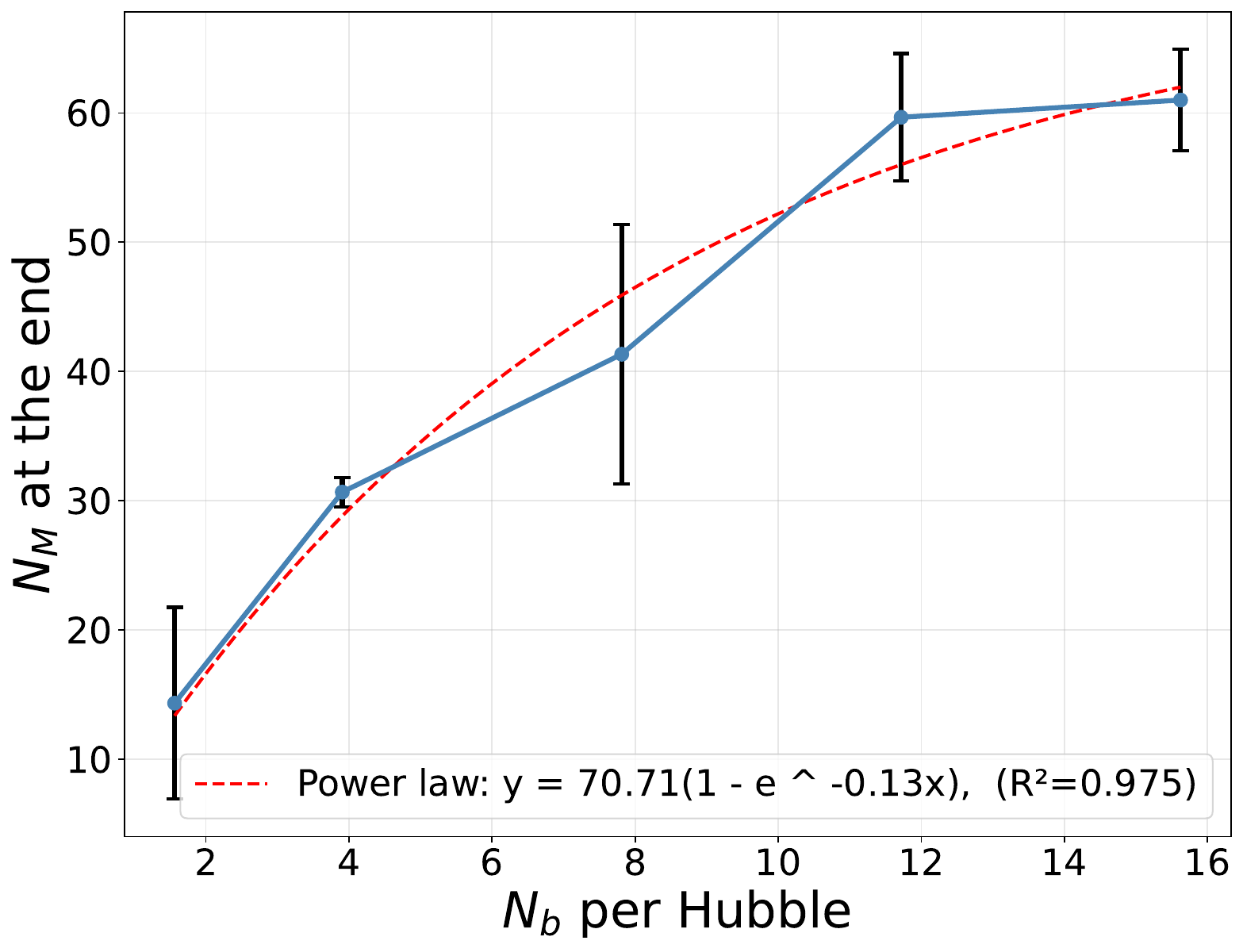}
    \caption{Top: Time evolution of the comoving monopole separation $d_{M}$ for various bubble numbers $N_{b}$, plotted from $\tau > 0$ immediately following the completion of the phase transition. 
    A higher initial bubble density results in a larger population of surviving monopole pairs. Bottom: Number of remaining monopole-antimonopole pairs at the end of the simulation as a function of $N_{b}$. The red dashed line denotes the theoretical fit, and the error bars represent the $95\%$ confidence intervals. }
    \label{fig:NM}
\end{figure}
Next, we explore the evolutionary dynamics of the monopole-antimonopole pairs produced during the phase transition. Crucially, since the system initially possesses no net magnetic charge, monopoles and antimonopoles are produced in exact symmetry, ensuring that the net topological charge remains strictly zero throughout the evolution. These topological defects are generated via bubble collisions and subsequently undergo pairwise annihilation once the PT completes. The top panel of Fig. \ref{fig:NM} illustrates the time evolution of the comoving monopole separation $d_{M}$ for various values of $N_{b}$. We observe that a larger $N_{b}$ yields a higher abundance of both produced and surviving monopoles, thus a shorter separation. 
With the exception of the $N_b = 100$ case, all other runs exhibit a similar growth in separation with time, signifying the scaling behavior.
This behavior occurs because closely separated pairs have already annihilated, whereas the attraction between the remaining widely spaced monopoles and antimonopoles is effectively counterbalanced by the cosmological expansion, halting further annihilation. Consequently, these stable relics can potentially persist into the present-day universe. It is worth noting that at the very beginning of the PRS stage, the number of monopoles exhibits a plateau. This is because we initially choose relatively large values of $\lambda$ and $g$, making the monopole size smaller than one lattice spacing $\Delta x$, which causes the monopoles to be ''pinned" to the lattice site and unable to move freely. As the PRS stage progresses, $\lambda$ and $g$ gradually decrease and the monopole size grows, allowing them to move within the box and annihilate due to mutual attraction, which leads to a decrease in the monopole count. 

To further quantify this relationship, we extract the population of surviving monopole-antimonopole pairs at the end of the simulation for different values of $N_{b}$, as presented in the bottom panel of Fig. \ref{fig:NM}. We find that as the number of bubbles increases, the monopoles exhibit a tendency toward saturation. We speculate that this is because below a certain distance, monopoles always annihilate very quickly. Furthermore, a larger number of bubbles leads to more densely generated monopoles. Due to their close proximity, these extra generated monopoles cannot survive for long. Therefore, we choose the following function for fitting:
\begin{align}
    N_M=70.71\left[1 - \exp (-0.13N_b)\right]\;.
\end{align}
The dependence of the monopole abundance on the bubble number is not particularly sensitive; as the number of bubbles per Hubble volume increases roughly from 1 to 10, the number of surviving monopole pairs rises from 10 to 60.

\subsection{Magnetic Fields and GWs Power Spectra}
\begin{figure}[!htp]
    \centering
    \includegraphics[width=0.8\linewidth]{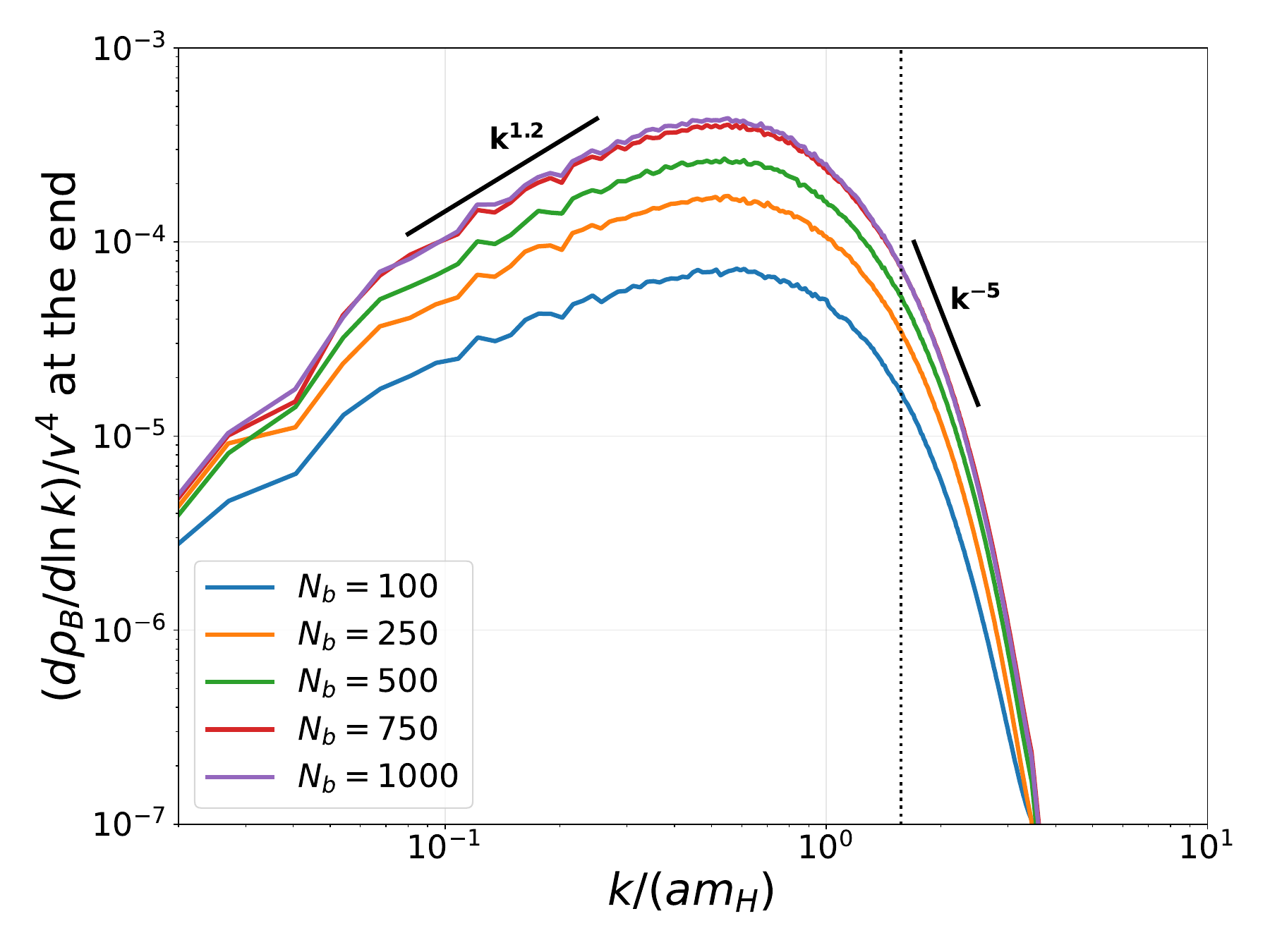}
    \caption{Magnetic field spectra at the end of the simulation for various values of $N_M$. The black dotted line denotes the physical size of the monopoles $k/a=2\pi/w_M$ at the final time step. 
    }
    \label{fig:B2final}
\end{figure}
In Fig.~\ref{fig:B2final}, we present the generated MF spectra for different bubble numbers $N_{b}$. We find that the magnetic field spectra corresponding to different monopole numbers exhibit qualitative similarities: they all feature a broken power-law profile with comparable peak frequencies. In the low-frequency (low-$k$) regime, the spectrum rises as $k^{-1.2}$. On scales smaller than the monopole size (dotted line in Fig.~\ref{fig:B2final}), the spectrum falls off rapidly with a slope of $k^{-5}$, demonstrating the magnetic screening effect inside the monopoles. Furthermore, as the number of monopoles increases, the peak energy density of the magnetic field also grows.

\begin{figure}
    \centering
    \includegraphics[width=0.8\linewidth]{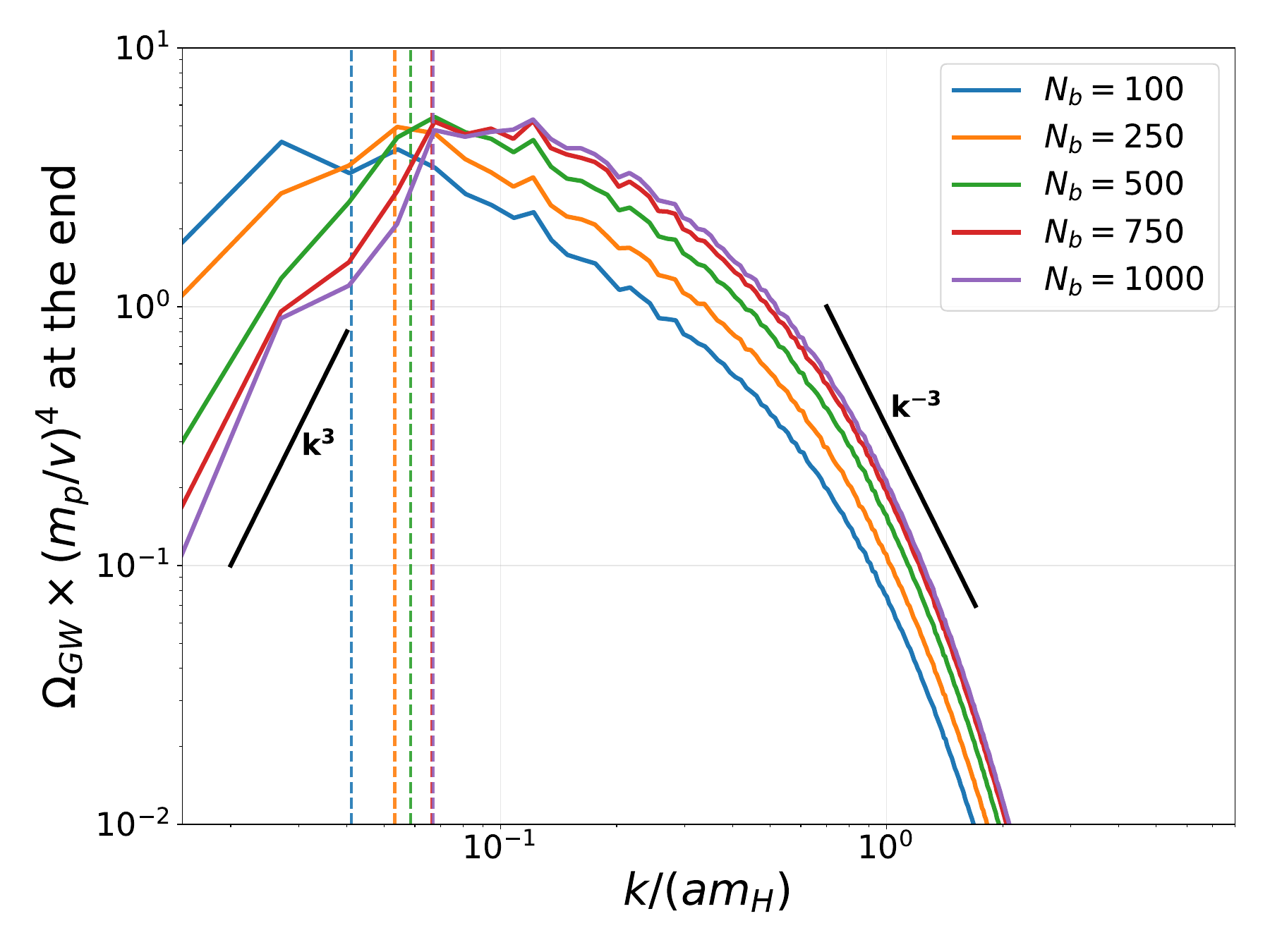}
    \caption{GW power spectra at the end of the simulation for various values of $N_M$. The physical scales corresponding to the monopole separation $k/a = 2\pi/d_M$ for different values of $N_M$ are indicated by dashed lines of the matching colors.
    }
    \label{fig:OmegaGWfinal}
\end{figure}
Subsequently, we investigate the GW signals generated by the monopole dynamics. Fig. \ref{fig:OmegaGWfinal} displays the time evolution of the GW power spectrum for various values of $N_{b}$. We observe that, first, across different monopole numbers, the power spectra similarly exhibit a broken power-law behavior: rising as $k^3$ in the low-frequency (low-$k$) regime (consistent with the predication in~\cite{Cai:2019cdl}), followed by a brief plateau, and falling off as $k^{-3}$ in the high-frequency (high-$k$) regime. Second, the peak frequency of the power spectrum consistently occurs at the scale of the mean monopole separation $d_M = L(2N_M)^{-1/3}$(dashed line in the figure). As the number of monopoles $N_M$ increases, the separation decreases, shifting the spectral peak to the right (toward higher frequencies). 

We also plot the time evolution of both the magnetic field and gravitational wave spectra for various $N_b$, which are provided in the \textit{Supplemental Material}. From these figures, the effects of the PRS method and the expansion on the spectral wavenumbers can be clearly observed.

\section{Cosmological constraints}


Since searches for magnetic monopoles to date have yielded null results, they must either be extremely scarce or constitute dark matter that participates in no interactions other than gravity. We discuss below the constraints on both the monopole mass
\begin{align}
    m_M = \frac{4\pi v}{g}f\left(\frac{m_H}{m_W}\right) = 43.71v\;,
\end{align}
and the phase transition temperature $T_*$ under each of these two scenarios. Here we take $m_H/m_W = 1.247$ and $f(1.247)\simeq1.273$~\cite{PhysRevD.71.119902}.

If magnetic monopoles are treated as visible matter, the most stringent limit arises from the Parker bound for nonrelativistic monopoles~\cite{MACRO:2002jdv,Parker:1970xv}
$
    F < 10^{-16}\,\mathrm{s}^{-1}\mathrm{cm}^{-2}\mathrm{sr}^{-1}\;$.
Assuming that monopoles move isotropically within intergalactic space or inside galaxies, and taking their typical velocity to be comparable to the galactic virial velocity $v_M  \sim 10^{-3}c  \approx 3 \times 10^7 \text{ cm/s}$, the contemporary monopole number density is given by \cite{Zhang_2024}:
\begin{align}
    n_{M,0} \lesssim \frac{4\pi F}{v_M} =  4.19 \times 10^{-23} \text{ cm}^{-3}\;.
\end{align}
Then the energy density of monopoles is 
\begin{align}
    \rho_{M,0} = n_{M,0}m_M\lesssim 1.83v\times 10^{-21} \text{ cm}^{-3}\;.
\end{align}
By substituting today's critical energy density $\rho_{c,0}/{h^2} \approx 1.05 \times 10^{-5} \text{ GeV/cm}^3$, we can calculate the present-day monopole abundance:
\begin{align}
    \Omega_{PB}h^2 = \frac{\rho_{M,0}}{\rho_{c,0}/{h^2}} \lesssim 1.74\times 10^{-16}\bigg(\frac{v}{1\,\mathrm{GeV}}\bigg)\;.\label{OmegaV}
\end{align}
Meanwhile, under the hypothesis that monopoles account for dark matter with the observational constraint: 
\begin{align}
    \Omega_{DM}h^2 \leqslant 0.12\;.\label{OmegaDM}
\end{align}

Next, we establish the relationship between the density parameter $\Omega_M h^2$ and the number of monopoles $N_M$. We assume the phase transition occurs at a critical temperature of $T_* \sim v$. Today's relic abundance of monopoles is given by:
\begin{align}
    \Omega_{M}h^2 &= \frac{Y_Ms_0m_M}{\rho_{c,0}/h^2} = 1.42\times10^{-45}N_M\bigg(\frac{v}{1\,\mathrm{GeV}}\bigg)^4 \nonumber\\
    &= 7.10\times10^{-44}\bigg(\frac{v}{1\,\mathrm{GeV}}\bigg)^4\;,\label{OmegaM}
\end{align}
where $Y_M = n_M/s = n_{M,0}/s_0$ is the monopole-to-entropy ratio, $s_0$ is the today entropy density. According to our simulations, we take the number of surviving monopole pairs as $N_M = 50$.

Finally, comparing Eqs. \eqref{OmegaV} and \eqref{OmegaM}, we find that under the assumption that monopoles constitute visible matter, their mass must satisfy $m_M \lesssim 2.18\times10^{11}$ GeV, with the corresponding phase transition temperature being approximately $T_* \lesssim 4.98\times10^9$ GeV. Comparing Eqs. \eqref{OmegaDM} and \eqref{OmegaM}, under the hypothesis that monopoles account for dark matter, their mass should be bounded by $m_M \lesssim 1.58\times10^{12}$ GeV, leading to a phase transition temperature of roughly $T_* \lesssim 3.60\times10^{10}$ GeV.

For a non-helical, inhomogeneous primordial magnetic field generated during the phase transition at temperature $T_*$, the field undergoes self-similar decaying magnetohydrodynamic (MHD) turbulence in the radiation-dominated plasma. Defining the comoving magnetic field as $\tilde{B} \equiv a^2 B$, the root-mean-square field strength decays as $\tilde{B}(\tau) \propto \tau^{-5/7}$ with conformal time $\tau$~\cite{Durrer_2013, Brandenburg_Durrer_Huang_Kahniashvili_Mandal_Mukohyama_2020}. This turbulent dissipation continues until recombination ($T_{\mathrm{rec}} \approx 0.3~\mathrm{eV}$), after which the drop in ionization fraction freezes the fluid velocity and the magnetic field redshifts purely adiabatically. The physical root-mean-square field strength today, $B_0$, is therefore related to the initial field strength $B_*$ by
\begin{align}
B_0 &= B_* \left( \frac{a_*}{a_0} \right)^2 \left( \frac{\tau_{\mathrm{rec}}}{\tau_*} \right)^{-5/7}\notag\\ 
&\simeq B_* \left( \frac{a_*}{a_0} \right)^2 \left[ \left( \frac{g_*(T_*)}{g_*(T_{\mathrm{rec}})} \right)^{1/6} \frac{T_*}{T_{\mathrm{rec}}} \right]^{-5/7}\;,
\end{align}
where $g_*(T_*)$ denotes the effective relativistic energy degrees of freedom. For temperatures well above the electroweak scale, $g_*(T_*)=106.75$. 
Substituting the results from Fig. \ref{fig:B2final} and adopting $v \sim T_* \sim 10^{10}\text{ GeV}$, the present-day magnetic field strength generated by monopoles is estimated to be $B_0 \sim 4.68\times10^{-23}\text{ Gauss}$, which falls below the lower bounds, $B_0 \sim 10^{-17}\text{ Gauss}$, reported by Fermi-LAT~\cite{Taylor_2011, Fermi-LAT:2018jdy}. This indicates that the origin of cosmological magnetic fields cannot be attributed to monopoles, pointing instead to alternative magnetogenesis mechanisms. 

The GW power spectrum generated during the radiation-dominated era, $\Omega_{\mathrm{GW},*}(f_*)$, is redshifted to the present day as the Universe expands. Because gravitational waves decouple immediately upon generation and propagate freely, their energy density scales as $\rho_{\mathrm{GW}} \propto a^{-4}$. Combining this with the conservation of entropy in the cosmic plasma, $s \propto g_{*s} a^3 T^3 = \mathrm{const}$, the present-day amplitude $\Omega_{\mathrm{GW},0}(f_0)h^2$ is related to the initial spectrum by
\begin{align}
&\ \Omega_{\mathrm{GW},0}(f_0) h^2 \notag\\
=&\ \Omega_{\mathrm{rad},0} h^2 \left( \frac{g_*(T_0)}{g_*(T_*)} \right) \left( \frac{g_{*s}(T_*)}{g_{*s}(T_0)} \right)^{4/3} \Omega_{\mathrm{GW},*}(f_*)\;,
\end{align}
where $\Omega_{\mathrm{rad},0} h^2 \approx 4.15 \times 10^{-5}$ represents the radiation relic density today, and $g_{*s}(T_*)$ denotes the effective relativistic entropy degrees of freedom at the transition temperature $T_*$. Assuming $g_*(T_*) = g_{*s}(T_*) \approx 106.75$ for temperatures well above the electroweak scale, this relation reduces to the standard parameterized form:
\begin{align}
\Omega_{\mathrm{GW},0}(f_0) h^2 \simeq 1.63 \times 10^{-5}  \Omega_{\mathrm{GW},*}(f_*)\;.
\end{align}
Substituting the numerical results from Fig. \ref{fig:OmegaGWfinal} and adopting $v \sim T_* \sim 10^{10}\text{ GeV}$ along with $m_\mathrm{p} = 2.44 \times 10^{18}\text{ GeV}$, the present-day peak amplitude of the gravitational waves produced by monopoles is approximately $\Omega_{\text{GW}} h^2 \sim 2.30\times10^{-38}$, which falls far below the sensitivity range of any current detector. This is consistent with the findings obtained in Ref.~\cite{Yang:2022quy}.

 
\section{ Conclusion and discussion}
In this work, we numerically study the monopole generation from a cosmological first-order phase transition on a 3D lattice, and derive the relation between the number density of magnetic monopole pairs and the vacuum bubbles, and the associated generation of the MF and GW spectra. The constraints on the monopole number density from the Parker bound and the relic abundance all preclude the detection of GW signatures. 


In order for the monopole abundance to satisfy current observational constraints, both its mass $m_M$ and the temperature $T_*$ of the first-order phase transition must be stringently constrained to values far below the Planck mass $m_\mathrm{p}$. Consequently, the gravitational waves generated by monopoles, when redshifted to the present day, would fall well below the sensitivity thresholds of OSQAR, ALP~\cite{Ejlli_2019}, ARCADE~\cite{Domcke_2021}, and GW-photon conversion in geomagnetic and galactic magnetic fields~\cite{Ito_2024}. 
Regarding the magnetic field, owing to the magnetohydrodynamic (MHD) dissipation of non-helical magnetic fields, a higher phase transition temperature leads to a weaker residual field strength today. According to our model, the magnetic field strength generated by monopoles also falls below the observational sensitivity $B_0 \sim 10^{-17}\text{ Gauss}$ established by Fermi-LAT~\cite{Taylor_2011, Fermi-LAT:2018jdy} observations of blazars. If we require the generated magnetic field to avoid being ruled out by this observational constraint, the phase transition temperature would need to satisfy $T_* \sim 350\text{ GeV}$.

 Regarding the dark matter relic abundance constraint, the relic density of monopoles produced via the Kibble-Zurek mechanism in a second-order phase transition incorporating monopole annihilation requires the phase transition temperature to satisfy $T_* < 10^6\sim10^9$ GeV~\cite{MURAYAMA2010162}, which is stricter than our bound. Ref. \cite{PhysRevLett.43.1365, bttg-zfr3, Brummer:2026wma} derived an analytical result incorporating monopole annihilation, yielding $Y_M = 2.64\times10^{-15}\propto T_*/m_\mathrm{p}$. In comparison, our numerical simulation gives a much smaller value, $Y_M = 7.80\times10^{-22}\propto (T_*/m_\mathrm{p})^3$. Therefore, our simulations show that, compared with their results, monopoles produced during the FOPT are far less likely to account for the full dark matter relic abundance.

We further note that, in the presence of monopoles, axions or axion-like particles acquire an effective mass via the Witten effect~\cite{Witten:1979ey}, thereby modifying predictions for the dark matter relic abundance~\cite{Fan:2021ntg,Kawasaki:2015lpf}. Ref.~\cite{Nomura:2015xil} propose to suppress axion isocurvature fluctuations with the hidden monopole.
It has also been observed that, if hidden monopoles constitute a fraction of dark matter, the effective mass can resolve the domain-wall problem~\cite{Nakagawa:2021nme,Sato:2018nqy,Zhou:2025zzz}. Moreover, the \(H_0\) tension may be alleviated through axion decay~\cite{Daido:2019tbm}.

\section{Acknowledgements}
We thank Ke-Pan Xie for helpful discussion.
 L.B. is supported by the National Natural Science Foundation of China (NSFC) under Grants Nos. 12322505, 12547101. L.B. also acknowledges Chongqing Natural Science Foundation under Grant No. CSTB2024NSCQ-JQX0022. R.G.C. is supported by the National Natural Science Foundation of China Grants Nos.12588101, 12235019. The numerical calculations in this study were carried out on the ORISE Supercomputer.

\bibliographystyle{apsrev}
\bibliography{references.bib}

@article{Hindmarsh:1993av,
    author = "Hindmarsh, Mark and Davis, Anne-Christine and Brandenberger, Robert H.",
    title = "{Formation of topological defects in first order phase transitions}",
    eprint = "hep-ph/9307203",
    archivePrefix = "arXiv",
    reportNumber = "DAMTP-93-13, BROWN-HET-902",
    doi = "10.1103/PhysRevD.49.1944",
    journal = "Phys. Rev. D",
    volume = "49",
    pages = "1944--1950",
    year = "1994"
}

@article{Kawasaki:2015lpf,
    author = "Kawasaki, Masahiro and Takahashi, Fuminobu and Yamada, Masaki",
    title = "{Suppressing the QCD Axion Abundance by Hidden Monopoles}",
    eprint = "1511.05030",
    archivePrefix = "arXiv",
    primaryClass = "hep-ph",
    reportNumber = "TU-1011, DESY-15-216",
    doi = "10.1016/j.physletb.2015.12.075",
    journal = "Phys. Lett. B",
    volume = "753",
    pages = "677--681",
    year = "2016"
}

@article{Nomura:2015xil,
    author = "Nomura, Yasunori and Rajendran, Surjeet and Sanches, Fabio",
    title = "{Axion Isocurvature and Magnetic Monopoles}",
    eprint = "1511.06347",
    archivePrefix = "arXiv",
    primaryClass = "hep-ph",
    reportNumber = "UCB-PTH-15-13",
    doi = "10.1103/PhysRevLett.116.141803",
    journal = "Phys. Rev. Lett.",
    volume = "116",
    number = "14",
    pages = "141803",
    year = "2016"
}

@article{Daido:2019tbm,
    author = "Daido, Ryuji and Ho, Shu-Yu and Takahashi, Fuminobu",
    title = "{Hidden monopole dark matter via axion portal and its implications for direct detection searches, beam-dump experiments, and the H$_{0}$ tension}",
    eprint = "1909.03627",
    archivePrefix = "arXiv",
    primaryClass = "hep-ph",
    reportNumber = "TU-1093, IPMU19-0122",
    doi = "10.1007/JHEP01(2020)185",
    journal = "JHEP",
    volume = "01",
    pages = "185",
    year = "2020"
}

@article{Zhou:2025zzz,
    author = "Zhou, Ruiyu and Bian, Ligong",
    title = "{Gravitational waves and dark matter with Witten effect}",
    eprint = "2501.09596",
    archivePrefix = "arXiv",
    primaryClass = "hep-ph",
    doi = "10.1088/1674-1137/adfc36",
    journal = "Chin. Phys. C",
    volume = "49",
    number = "11",
    pages = "115110",
    year = "2025"
}

@article{Sato:2018nqy,
    author = "Sato, Ryosuke and Takahashi, Fuminobu and Yamada, Masaki",
    title = "{Unified Origin of Axion and Monopole Dark Matter, and Solution to the Domain-wall Problem}",
    eprint = "1805.10533",
    archivePrefix = "arXiv",
    primaryClass = "hep-ph",
    reportNumber = "TU-1062; IPMU18-0079; MIT-CTP/5017, TU-1062, IPMU18-0079, MIT-CTP-5017",
    doi = "10.1103/PhysRevD.98.043535",
    journal = "Phys. Rev. D",
    volume = "98",
    number = "4",
    pages = "043535",
    year = "2018"
}

@article{Nakagawa:2021nme,
    author = "Nakagawa, Shota and Takahashi, Fuminobu and Yamada, Masaki",
    title = "{Cosmic Birefringence Triggered by Dark Matter Domination}",
    eprint = "2103.08153",
    archivePrefix = "arXiv",
    primaryClass = "hep-ph",
    reportNumber = "TU-1118, IPMU21-0018",
    doi = "10.1103/PhysRevLett.127.181103",
    journal = "Phys. Rev. Lett.",
    volume = "127",
    number = "18",
    pages = "181103",
    year = "2021"
}

@article{Fan:2021ntg,
    author = "Fan, JiJi and Fraser, Katherine and Reece, Matthew and Stout, John",
    title = "{Axion Mass from Magnetic Monopole Loops}",
    eprint = "2105.09950",
    archivePrefix = "arXiv",
    primaryClass = "hep-ph",
    doi = "10.1103/PhysRevLett.127.131602",
    journal = "Phys. Rev. Lett.",
    volume = "127",
    number = "13",
    pages = "131602",
    year = "2021"
}

@article{Witten:1979ey,
    author = "Witten, Edward",
    title = "{Dyons of Charge e theta/2 pi}",
    reportNumber = "CERN-TH-2724",
    doi = "10.1016/0370-2693(79)90838-4",
    journal = "Phys. Lett. B",
    volume = "86",
    pages = "283--287",
    year = "1979"
}

@article{Yang:2022quy,
    author = "Yang, Jing and Zhou, Ruiyu and Bian, Ligong",
    title = "{Gravitational waves and monopoles dark matter from first-order phase transition}",
    eprint = "2204.07540",
    archivePrefix = "arXiv",
    primaryClass = "hep-ph",
    doi = "10.1016/j.physletb.2023.137822",
    journal = "Phys. Lett. B",
    volume = "839",
    pages = "137822",
    year = "2023"
}

@article{Guth:1979bh,
    author = "Guth, Alan H. and Tye, S. H. H.",
    title = "{Phase Transitions and Magnetic Monopole Production in the Very Early Universe}",
    reportNumber = "SLAC-PUB-2448, CLNS-79-441",
    doi = "10.1103/PhysRevLett.44.631",
    journal = "Phys. Rev. Lett.",
    volume = "44",
    pages = "631",
    year = "1980",
    note = "[Erratum: Phys.Rev.Lett. 44, 963 (1980)]"
}

@article{Kibble:1995aa,
    author = "Kibble, T. W. B. and Vilenkin, Alexander",
    title = "{Phase equilibration in bubble collisions}",
    eprint = "hep-ph/9501266",
    archivePrefix = "arXiv",
    reportNumber = "IMPERIAL-TP-94-95-11, TUTP-95-2, NI-94039, NI94039",
    doi = "10.1103/PhysRevD.52.679",
    journal = "Phys. Rev. D",
    volume = "52",
    pages = "679--688",
    year = "1995"
}

@article{Copeland:1996jz,
    author = "Copeland, Edmund J. and Saffin, P. M.",
    title = "{Bubble collisions in Abelian gauge theories and the geodesic rule}",
    eprint = "hep-ph/9604231",
    archivePrefix = "arXiv",
    reportNumber = "SUSX-TH-96-006",
    doi = "10.1103/PhysRevD.54.6088",
    journal = "Phys. Rev. D",
    volume = "54",
    pages = "6088--6094",
    year = "1996"
}

@article{Creaco:2008ey,
    author = "Creaco, Anthony J. and Kalogeropoulos, Nikos",
    title = "{The Geodesic rule for higher codimensional global defects}",
    eprint = "0805.2353",
    archivePrefix = "arXiv",
    primaryClass = "hep-th",
    doi = "10.1142/S0217732308027242",
    journal = "Mod. Phys. Lett. A",
    volume = "23",
    pages = "2053--2066",
    year = "2008"
}

@article{Georgi:1974sy,
    author = "Georgi, H. and Glashow, S. L.",
    title = "{Unity of All Elementary Particle Forces}",
    doi = "10.1103/PhysRevLett.32.438",
    journal = "Phys. Rev. Lett.",
    volume = "32",
    pages = "438--441",
    year = "1974"
}

@article{Kibble_1976,
doi = {10.1088/0305-4470/9/8/029},
url = {https://doi.org/10.1088/0305-4470/9/8/029},
year = {1976},
month = {aug},
publisher = {},
volume = {9},
number = {8},
pages = {1387},
author = {T W B Kibble},
title = {Topology of cosmic domains and strings},
journal = {Journal of Physics A: Mathematical and General}
}

@article{Ejlli_2019,
   title={Upper limits on the amplitude of ultra-high-frequency gravitational waves from graviton to photon conversion},
   volume={79},
   ISSN={1434-6052},
   url={http://dx.doi.org/10.1140/epjc/s10052-019-7542-5},
   DOI={10.1140/epjc/s10052-019-7542-5},
   number={12},
   journal={The European Physical Journal C},
   publisher={Springer Science and Business Media LLC},
   author={Ejlli, A. and Ejlli, D. and Cruise, A. M. and Pisano, G. and Grote, H.},
   year={2019},
   month=Dec }

@article{Domcke_2021,
   title={Potential of Radio Telescopes as High-Frequency Gravitational Wave Detectors},
   volume={126},
   ISSN={1079-7114},
   url={http://dx.doi.org/10.1103/PhysRevLett.126.021104},
   DOI={10.1103/physrevlett.126.021104},
   number={2},
   journal={Physical Review Letters},
   publisher={American Physical Society (APS)},
   author={Domcke, Valerie and Garcia-Cely, Camilo},
   year={2021},
   month=Jan }

@article{Ito_2024,
   title={Gravitational Wave Search through Electromagnetic Telescopes},
   volume={2024},
   ISSN={2050-3911},
   url={http://dx.doi.org/10.1093/ptep/ptae004},
   DOI={10.1093/ptep/ptae004},
   number={2},
   journal={Progress of Theoretical and Experimental Physics},
   publisher={Oxford University Press (OUP)},
   author={Ito, Asuka and Kohri, Kazunori and Nakayama, Kazunori},
   year={2024},
   month=Jan }

@article{Kibble:1980mv,
    author = "Kibble, T. W. B.",
    title = "{Some Implications of a Cosmological Phase Transition}",
    reportNumber = "ICTP-79-80-23",
    doi = "10.1016/0370-1573(80)90091-5",
    journal = "Phys. Rept.",
    volume = "67",
    pages = "183",
    year = "1980"
}

@article{tHooft:1974kcl,
    author = "'t Hooft, Gerard",
    editor = "Taylor, J. C.",
    title = "{Magnetic Monopoles in Unified Gauge Theories}",
    reportNumber = "CERN-TH-1876",
    doi = "10.1016/0550-3213(74)90486-6",
    journal = "Nucl. Phys. B",
    volume = "79",
    pages = "276--284",
    year = "1974"
}

@article{Polyakov:1974ek,
    author = "Polyakov, Alexander M.",
    editor = "Taylor, J. C.",
    title = "{Particle Spectrum in Quantum Field Theory}",
    reportNumber = "PRINT-74-1566 (LANDAU-INST)",
    journal = "JETP Lett.",
    volume = "20",
    pages = "194--195",
    year = "1974"
}

@article{Preskill:1979zi,
    author = "Preskill, John",
    title = "{Cosmological Production of Superheavy Magnetic Monopoles}",
    reportNumber = "HUTP-79/A028",
    doi = "10.1103/PhysRevLett.43.1365",
    journal = "Phys. Rev. Lett.",
    volume = "43",
    pages = "1365",
    year = "1979"
}

@article{Linde:1981mu,
    author = "Linde, Andrei D.",
    editor = "Fang, Li-Zhi and Ruffini, R.",
    title = "{A New Inflationary Universe Scenario: A Possible Solution of the Horizon, Flatness, Homogeneity, Isotropy and Primordial Monopole Problems}",
    reportNumber = "LEBEDEV-81-229",
    doi = "10.1016/0370-2693(82)91219-9",
    journal = "Phys. Lett. B",
    volume = "108",
    pages = "389--393",
    year = "1982"
}

@article{Coleman:1977py,
    author = "Coleman, Sidney R.",
    title = "{The Fate of the False Vacuum. 1. Semiclassical Theory}",
    reportNumber = "HUTP-77-A004",
    doi = "10.1103/PhysRevD.15.2929",
    journal = "Phys. Rev. D",
    volume = "15",
    pages = "2929--2936",
    year = "1977",
    note = "[Erratum: Phys.Rev.D 16, 1248 (1977)]"
}

@article{Linde:1981zj,
    author = "Linde, Andrei D.",
    title = "{Decay of the False Vacuum at Finite Temperature}",
    reportNumber = "LEBEDEV-81-265",
    doi = "10.1016/0550-3213(83)90072-X",
    journal = "Nucl. Phys. B",
    volume = "216",
    pages = "421",
    year = "1983",
    note = "[Erratum: Nucl.Phys.B 223, 544 (1983)]"
}

@article{Guth:1981uk,
    author = "Guth, Alan H. and Weinberg, Erick J.",
    title = "{Cosmological Consequences of a First Order Phase Transition in the SU(5) Grand Unified Model}",
    reportNumber = "CU-TP-183",
    doi = "10.1103/PhysRevD.23.876",
    journal = "Phys. Rev. D",
    volume = "23",
    pages = "876",
    year = "1981"
}

@article{Di:2025ncl,
    author = "Di, Yuefeng and Bian, Ligong and Cai, Rong-Gen",
    title = "{Impact of primordial magnetic fields on the first-order electroweak phase transition}",
    eprint = "2508.07416",
    archivePrefix = "arXiv",
    primaryClass = "hep-ph",
    doi = "10.1103/jrd3-s1h2",
    journal = "Phys. Rev. D",
    volume = "113",
    number = "4",
    pages = "043529",
    year = "2026"
}

@article{Di:2024gsl,
    author = "Di, Yuefeng and Bian, Ligong and Cai, Rong-Gen",
    title = "{Baryogenesis Induced by Magnetic Field Effects During the Electroweak Phase Transition}",
    eprint = "2409.16124",
    archivePrefix = "arXiv",
    primaryClass = "hep-ph",
    month = "9",
    year = "2024"
}

@article{Caldwell:2022qsj,
    author = "Caldwell, Robert and others",
    title = "{Detection of early-universe gravitational-wave signatures and fundamental physics}",
    eprint = "2203.07972",
    archivePrefix = "arXiv",
    primaryClass = "gr-qc",
    doi = "10.1007/s10714-022-03027-x",
    journal = "Gen. Rel. Grav.",
    volume = "54",
    number = "12",
    pages = "156",
    year = "2022"
}

@article{Roshan:2024qnv,
    author = "Roshan, Rishav and White, Graham",
    title = "{Using gravitational waves to see the first second of the Universe}",
    eprint = "2401.04388",
    archivePrefix = "arXiv",
    primaryClass = "hep-ph",
    doi = "10.1103/RevModPhys.97.015001",
    journal = "Rev. Mod. Phys.",
    volume = "97",
    number = "1",
    pages = "015001",
    year = "2025"
}

@article{Bevis_2010,
   title={CMB power spectra from cosmic strings: Predictions for the Planck satellite and beyond},
   volume={82},
   ISSN={1550-2368},
   url={http://dx.doi.org/10.1103/PhysRevD.82.065004},
   DOI={10.1103/physrevd.82.065004},
   number={6},
   journal={Physical Review D},
   publisher={American Physical Society (APS)},
   author={Bevis, Neil and Hindmarsh, Mark and Kunz, Martin and Urrestilla, Jon},
   year={2010},
   month=Sept }

@ARTICLE{1989ApJ...347..590P,
       author = {{Press}, William H. and {Ryden}, Barbara S. and {Spergel}, David N.},
        title = "{Dynamical Evolution of Domain Walls in an Expanding Universe}",
      journal = {\apj},
         year = 1989,
        month = dec,
       volume = {347},
        pages = {590},
          doi = {10.1086/168151},
       adsurl = {https://ui.adsabs.harvard.edu/abs/1989ApJ...347..590P}
}

@article{Cai:2019cdl,
    author = "Cai, Rong-Gen and Pi, Shi and Sasaki, Misao",
    title = "{Universal infrared scaling of gravitational wave background spectra}",
    eprint = "1909.13728",
    archivePrefix = "arXiv",
    primaryClass = "astro-ph.CO",
    reportNumber = "IPMU19-0135, YITP-19-88",
    doi = "10.1103/PhysRevD.102.083528",
    journal = "Phys. Rev. D",
    volume = "102",
    number = "8",
    pages = "083528",
    year = "2020"
}

@ARTICLE{crank1996,
       author = {Crank, J. and Nicolson, P.},
        title = "{A practical method for numerical evaluation of solutions of partial differential equations of the heat-conduction type}",
      journal = {Adv Comput Math},
         year = 1996,
        month = dec,
       volume = {6},
        pages = {207–226},
          doi = {10.1007/BF02127704},
       adsurl = {https://doi.org/10.1007/BF02127704},
}

@article{Hindmarsh:2025vxh,
    author = {Hindmarsh, Mark and Lopez-Eiguren, Asier and Sepp{\"a}, Riikka and Weir, David J.},
    title = "{Numerical simulations of magnetic monopole evolution in an expanding universe}",
    eprint = "2511.14204",
    archivePrefix = "arXiv",
    primaryClass = "astro-ph.CO",
    reportNumber = "HIP-2025-31/TH",
    month = "11",
    year = "2025"
}

@article{MURAYAMA2010162,
title = {Topological dark matter},
journal = {Physics Letters B},
volume = {686},
number = {2},
pages = {162-165},
year = {2010},
issn = {0370-2693},
doi = {https://doi.org/10.1016/j.physletb.2010.02.037},
url = {https://www.sciencedirect.com/science/article/pii/S0370269310002248},
author = {Hitoshi Murayama and Jing Shu}
}

@article{PhysRevLett.43.1365,
  title = {Cosmological Production of Superheavy Magnetic Monopoles},
  author = {Preskill, John P.},
  journal = {Phys. Rev. Lett.},
  volume = {43},
  issue = {19},
  pages = {1365--1368},
  numpages = {0},
  year = {1979},
  month = {Nov},
  publisher = {American Physical Society},
  doi = {10.1103/PhysRevLett.43.1365},
  url = {https://link.aps.org/doi/10.1103/PhysRevLett.43.1365}
}

@article{bttg-zfr3,
  title = {No room for minimal monopole dark matter},
  author = {Br{\"u}mmer, Felix and Ferrante, Giacomo and Fischer, Th{\'e}odore  and Frigerio, Michele},
  journal = {Phys. Rev. D},
  volume = {113},
  issue = {9},
  pages = {L091701},
  numpages = {7},
  year = {2026},
  month = {May},
  publisher = {American Physical Society},
  doi = {10.1103/bttg-zfr3},
  url = {https://link.aps.org/doi/10.1103/bttg-zfr3}
}

@article{Brummer:2026wma,
    author = {Br{\"u}mmer, Felix and Ferrante, Giacomo and Fischer, Th{\'e}odore and Frigerio, Michele},
    title = "{The price for monopole dark matter}",
    eprint = "2607.29492",
    archivePrefix = "arXiv",
    primaryClass = "hep-ph",
    reportNumber = "ULB-TH/26-22",
    month = "7",
    year = "2026"
}

@article{PhysRevD.71.119902,
  title = {Publisher's Note: Numerical and asymptotic analysis of the 't Hooft-Polyakov magnetic monopole [Phys. Rev. D 71, 035002 (2005)]},
  author = {Forg{\'a}cs, P. and Obadia, N. and Reuillon, S.},
  journal = {Phys. Rev. D},
  volume = {71},
  issue = {11},
  pages = {119902},
  numpages = {1},
  year = {2005},
  month = {Jun},
  publisher = {American Physical Society},
  doi = {10.1103/PhysRevD.71.119902},
  url = {https://link.aps.org/doi/10.1103/PhysRevD.71.119902}
}

@article{MACRO:2002jdv,
    author = "Ambrosio, M. and others",
    collaboration = "MACRO",
    title = "{Final results of magnetic monopole searches with the MACRO experiment}",
    eprint = "hep-ex/0207020",
    archivePrefix = "arXiv",
    reportNumber = "DFUB-02-05",
    doi = "10.1140/epjc/s2002-01046-9",
    journal = "Eur. Phys. J. C",
    volume = "25",
    pages = "511--522",
    year = "2002"
}

@article{Zhang_2024,
   title={On the cosmological abundance of magnetic monopoles},
   volume={2024},
   ISSN={1029-8479},
   url={http://dx.doi.org/10.1007/JHEP08(2024)220},
   DOI={10.1007/jhep08(2024)220},
   number={8},
   journal={Journal of High Energy Physics},
   publisher={Springer Science and Business Media LLC},
   author={Zhang, Chen and Zhang, Shi-Hao and Fu, Bowen and Zhang, Jing-Fei and Zhang, Xin},
   year={2024},
   month=Aug }

@article{Hindmarsh_Rummukainen_Weir_2017,  
 title={Numerical simulations of necklaces in SU(2) gauge-Higgs field theory}, 
 url={http://dx.doi.org/10.1103/physrevd.95.063520}, 
 DOI={10.1103/physrevd.95.063520}, 
 journal={Physical Review D}, 
 author={Hindmarsh, Mark and Rummukainen, Kari and Weir, David J.}, 
 year={2017}, 
 month={Mar}, 
 language={en-US} 
 }

@article{Edwards_2009,
   title={’t Hooft-Polyakov monopoles in lattice<mml:math xmlns:mml=“http://www.w3.org/1998/Math/MathML” display=“inline”><mml:mi>SU</mml:mi><mml:mo stretchy=“false”>(</mml:mo><mml:mi>N</mml:mi><mml:mo stretchy=“false”>)</mml:mo><mml:mo>+</mml:mo><mml:mtext mathvariant=“normal”>adjoint Higgs</mml:mtext></mml:math>theory},
   volume={80},
   ISSN={1550-2368},
   url={http://dx.doi.org/10.1103/PhysRevD.80.065030},
   DOI={10.1103/physrevd.80.065030},
   number={6},
   journal={Physical Review D},
   publisher={American Physical Society (APS)},
   author={Edwards, S. and Mehta, D. and Rajantie, A. and von Smekal, L.},
   year={2009},
   month=Sept }

@article{Grasso:1997nx,
    author = "Grasso, Dario and Riotto, Antonio",
    title = "{On the nature of the magnetic fields generated during the electroweak phase transition}",
    eprint = "hep-ph/9707265",
    archivePrefix = "arXiv",
    reportNumber = "FERMILAB-PUB-97-209-A",
    doi = "10.1016/S0370-2693(97)01224-0",
    journal = "Phys. Lett. B",
    volume = "418",
    pages = "258--265",
    year = "1998"
}

@article{Enqvist:1993np,
    author = "Enqvist, K. and Olesen, P.",
    title = "{On primordial magnetic fields of electroweak origin}",
    eprint = "hep-ph/9308270",
    archivePrefix = "arXiv",
    reportNumber = "NBI-HE-93-33",
    doi = "10.1016/0370-2693(93)90799-N",
    journal = "Phys. Lett. B",
    volume = "319",
    pages = "178--185",
    year = "1993"
}

@article{Baym:1995fk,
    author = "Baym, Gordon and Bodeker, Dietrich and McLerran, Larry D.",
    title = "{Magnetic fields produced by phase transition bubbles in the electroweak phase transition}",
    eprint = "hep-ph/9507429",
    archivePrefix = "arXiv",
    reportNumber = "TPI-MINN-95-19-T, NUC-MINN-95-15-T, HEP-MINN-95-1344",
    doi = "10.1103/PhysRevD.53.662",
    journal = "Phys. Rev. D",
    volume = "53",
    pages = "662--667",
    year = "1996"
}

@article{Vachaspati:1991nm,
    author = "Vachaspati, T.",
    title = "{Magnetic fields from cosmological phase transitions}",
    doi = "10.1016/0370-2693(91)90051-Q",
    journal = "Phys. Lett. B",
    volume = "265",
    pages = "258--261",
    year = "1991"
}

@article{Durrer_2013,
   title={Cosmological magnetic fields: their generation, evolution and observation},
   volume={21},
   ISSN={1432-0754},
   url={http://dx.doi.org/10.1007/s00159-013-0062-7},
   DOI={10.1007/s00159-013-0062-7},
   number={1},
   journal={The Astronomy and Astrophysics Review},
   publisher={Springer Science and Business Media LLC},
   author={Durrer, Ruth and Neronov, Andrii},
   year={2013},
   month=jun }

@article{Fermi-LAT:2018jdy,
    author = "Ackermann, M. and others",
    collaboration = "Fermi-LAT",
    title = "{The Search for Spatial Extension in High-latitude Sources Detected by the $Fermi$ Large Area Telescope}",
    eprint = "1804.08035",
    archivePrefix = "arXiv",
    primaryClass = "astro-ph.HE",
    doi = "10.3847/1538-4365/aacdf7",
    journal = "Astrophys. J. Suppl.",
    volume = "237",
    number = "2",
    pages = "32",
    year = "2018"
}

@article{Guada_2020,
   title={FindBounce: Package for multi-field bounce actions},
   volume={256},
   ISSN={0010-4655},
   url={http://dx.doi.org/10.1016/j.cpc.2020.107480},
   DOI={10.1016/j.cpc.2020.107480},
   journal={Computer Physics Communications},
   publisher={Elsevier BV},
   author={Guada, Victor and Nemevšek, Miha and Pintar, Matevž},
   year={2020},
   month=nov, pages={107480} }

@article{Taylor_2011,
   title={Extragalactic magnetic fields constraints from simultaneous GeV–TeV observations of blazars},
   volume={529},
   ISSN={1432-0746},
   url={http://dx.doi.org/10.1051/0004-6361/201116441},
   DOI={10.1051/0004-6361/201116441},
   journal={Astronomy amp; Astrophysics},
   publisher={EDP Sciences},
   author={Taylor, A. M. and Vovk, I. and Neronov, A.},
   year={2011},
   month=apr, pages={A144} }

@article{Mazumdar_2019,
   title={Review of cosmic phase transitions: their significance and experimental signatures},
   volume={82},
   ISSN={1361-6633},
   url={http://dx.doi.org/10.1088/1361-6633/ab1f55},
   DOI={10.1088/1361-6633/ab1f55},
   number={7},
   journal={Reports on Progress in Physics},
   publisher={IOP Publishing},
   author={Mazumdar, Anupam and White, Graham},
   year={2019},
   month=jun, pages={076901} }

@article{Hindmarsh_2021,
   title={Phase transitions in the early universe},
   ISSN={2590-1990},
   url={http://dx.doi.org/10.21468/SciPostPhysLectNotes.24},
   DOI={10.21468/scipostphyslectnotes.24},
   journal={SciPost Physics Lecture Notes},
   publisher={Stichting SciPost},
   author={Hindmarsh, Mark and Lüben, Marvin and Lumma, Johannes and Pauly, Martin},
   year={2021},
   month=feb }

@article{Athron_2024,
   title={Cosmological phase transitions: From perturbative particle physics to gravitational waves},
   volume={135},
   ISSN={0146-6410},
   url={http://dx.doi.org/10.1016/j.ppnp.2023.104094},
   DOI={10.1016/j.ppnp.2023.104094},
   journal={Progress in Particle and Nuclear Physics},
   publisher={Elsevier BV},
   author={Athron, Peter and Balázs, Csaba and Fowlie, Andrew and Morris, Lachlan and Wu, Lei},
   year={2024},
   month=feb, pages={104094} }

@article{Di_2021,
   title={Magnetic Field and Gravitational Waves from the First-Order Phase Transition},
   volume={126},
   ISSN={1079-7114},
   url={http://dx.doi.org/10.1103/PhysRevLett.126.251102},
   DOI={10.1103/physrevlett.126.251102},
   number={25},
   journal={Physical Review Letters},
   publisher={American Physical Society (APS)},
   author={Di, Yuefeng and Wang, Jialong and Zhou, Ruiyu and Bian, Ligong and Cai, Rong-Gen and Liu, Jing},
   year={2021},
   month=jun }

@article{Yang_2022,
   title={Magnetic field generation from bubble collisions during first-order phase transition},
   volume={106},
   ISSN={2470-0029},
   url={http://dx.doi.org/10.1103/PhysRevD.106.023510},
   DOI={10.1103/physrevd.106.023510},
   number={2},
   journal={Physical Review D},
   publisher={American Physical Society (APS)},
   author={Yang, Jing and Bian, Ligong},
   year={2022},
   month=jul }

@article{PhysRevLett.51.1488,
  title = {Magnetohydrodynamic Effects of a First-Order Cosmological Phase Transition},
  author = {Hogan, Craig J.},
  journal = {Phys. Rev. Lett.},
  volume = {51},
  issue = {16},
  pages = {1488--1491},
  numpages = {0},
  year = {1983},
  month = {Oct},
  publisher = {American Physical Society},
  doi = {10.1103/PhysRevLett.51.1488},
  url = {https://link.aps.org/doi/10.1103/PhysRevLett.51.1488}
}

@article{Caprini_2016,
   title={Science with the space-based interferometer eLISA. II: gravitational waves from cosmological phase transitions},
   volume={2016},
   ISSN={1475-7516},
   url={http://dx.doi.org/10.1088/1475-7516/2016/04/001},
   DOI={10.1088/1475-7516/2016/04/001},
   number={04},
   journal={Journal of Cosmology and Astroparticle Physics},
   publisher={IOP Publishing},
   author={Caprini, Chiara and others},
   year={2016},
   month=apr, pages={001–001} }

@article{Caprini_2020,
   title={Detecting gravitational waves from cosmological phase transitions with LISA: an update},
   volume={2020},
   ISSN={1475-7516},
   url={http://dx.doi.org/10.1088/1475-7516/2020/03/024},
   DOI={10.1088/1475-7516/2020/03/024},
   number={03},
   journal={Journal of Cosmology and Astroparticle Physics},
   publisher={IOP Publishing},
   author={Caprini, Chiara and others},
   year={2020},
   month=mar, pages={024–024} }

@article{Brandenburg_Durrer_Huang_Kahniashvili_Mandal_Mukohyama_2020,  
 title={Primordial magnetic helicity evolution with a homogeneous magnetic field from inflation}, 
 url={http://dx.doi.org/10.1103/physrevd.102.023536}, 
 DOI={10.1103/physrevd.102.023536}, 
 journal={Physical Review D}, 
 author={Brandenburg, Axel and Durrer, Ruth and Huang, Yiwen and Kahniashvili, Tina and Mandal, Sayan and Mukohyama, Shinji}, 
 year={2020}, 
 month={Jul}, 
 language={en-US} 
 }

@article{fv7q-vsbn,
  title = {Standard Model Baryon Number Violation at Zero Temperature from Higgs Bubble Collisions},
  author = {Bhusal, Nabeen and Blasi, Simone and Cataldi, Martina and Chatrchyan, Aleksandr and Gorghetto, Marco and Servant, G\'eraldine},
  journal = {Phys. Rev. Lett.},
  volume = {136},
  issue = {24},
  pages = {241803},
  numpages = {10},
  year = {2026},
  month = {Jun},
  publisher = {American Physical Society},
  doi = {10.1103/fv7q-vsbn},
  url = {https://link.aps.org/doi/10.1103/fv7q-vsbn}
}

@article{Parker:1970xv,
    author = "Parker, Eugene N.",
    title = "{The Origin of Magnetic Fields}",
    reportNumber = "RX-977",
    doi = "10.1086/150442",
    journal = "Astrophys. J.",
    volume = "160",
    pages = "383",
    year = "1970"
}

@article{Davis_2000,
doi = {10.1088/1126-6708/2000/11/010},
url = {https://doi.org/10.1088/1126-6708/2000/11/010},
year = {2000},
month = {dec},
publisher = {},
volume = {2000},
number = {11},
pages = {010},
author = {Anne C. Davis and Tom W.B. Kibble and Arttu Rajantie and Hugh P. Shanahan},
title = {Topological defects in lattice gauge theories},
journal = {Journal of High Energy Physics}
}

@article{HOOFT1974276,
title = {Magnetic monopoles in unified gauge theories},
journal = {Nuclear Physics B},
volume = {79},
number = {2},
pages = {276-284},
year = {1974},
issn = {0550-3213},
doi = {https://doi.org/10.1016/0550-3213(74)90486-6},
url = {https://www.sciencedirect.com/science/article/pii/0550321374904866},
author = {G.'t Hooft}
}

@article{Tranberg:2017lrx,
    author = {Tranberg, Anders and T{\"a}htinen, Sara and Weir, David J.},
    title = "{Gravitational waves from non-Abelian gauge fields at a tachyonic transition}",
    eprint = "1706.02365",
    archivePrefix = "arXiv",
    primaryClass = "hep-ph",
    reportNumber = "HIP-2017-09-TH",
    doi = "10.1088/1475-7516/2018/04/012",
    journal = "JCAP",
    volume = "04",
    pages = "012",
    year = "2018"
}
\newpage

\onecolumngrid
\appendix

\section{Equations of motion on lattice}
Firstly, we introduce the link fields $U_i$ to describe the gauge fields on the lattice
\begin{align}
    U_i(\tau,x) = \exp\left[-\ii g\Delta xW_i(\tau,x)\right] = \exp\left[-\ii g\Delta x\frac{\sigma^a}{2}W_i^a(\tau,x)\right]\;.\label{Ui}
\end{align}
Since we adopt the temporal gauge, we have $W_0=0$. 
The spatial covariant derivatives are
\begin{align}
    D_i\phi(\tau,x) &= \frac{1}{\Delta x}[U_i(\tau,x)\phi(\tau,x+i)U_i^\dagger(\tau,x) - \phi(\tau,x)]\;.
\end{align}

We define $\pm 0$ to denote a forward or backward step $\Delta \tau$ in the temporal direction, and $\pm i$ to denote a forward or backward step $\Delta x$ along the spatial $i$-direction. We introduce the conjugate momentum fields $\pi$ and $F_i$ corresponding to $\phi$ and $U_i$, respectively. Temporally, $\phi$ and $U_i$ are defined at integer time steps, whereas $\pi$ and $F_i$ are defined at half-integer time steps. Spatially, $\phi$ and $\pi$ reside on the lattice sites, while $U_i$ and $F_i$ are defined on the links between adjacent sites.

Prior to evolving the equations of motion, we must address the time dependence of the gauge coupling $g$ introduced by the PRS prescription. As evident from Eq. \eqref{Ui}, $g$ enters directly into the definition of the link variable $U_i$, as well as the conjugate momenta $F_i$. Consequently, at the beginning of each time step, we first ``align" the coupling parameter via: 
\begin{align}
    U_{i,\,\mathrm{new}}(\tau,x) &= -\ii\frac{\sigma^a}{2}\frac{g(\tau)}{g(\tau-0)}\tr[\ii \sigma^aU_{i,\,\mathrm{old}}(\tau,x)]\;,\\
    F_{i,\,\mathrm{new}}(\tau,x) &= -\ii\frac{\sigma^a}{2}\frac{g(\tau)}{g(\tau-0)}\tr[\ii \sigma^aF_{i,\,\mathrm{old}}(\tau,x)]\;.
\end{align}
The time evolution within the current time step is then carried out using $U_{i,\,\mathrm{new}}$ and $F_{i,\,\mathrm{new}}(\tau,x)$. For notational convenience, we drop the subscript 'new' hereafter. Regarding the coupling parameter $\lambda$, because it enters exclusively through the potential term, no special treatment is required. 
Then, employing the leapfrog algorithm, we first update $\phi$ and $U_i$ using the following equations:
\begin{align}
    \phi(\tau+0,x)&=\phi(\tau,x)+a^{-2}\pi(\tau,x)\Delta \tau\;,\label{eomphi}\\
    U_i(\tau+0,x)&=g(\tau)\Delta\tau\Delta xF_i(\tau,x)U_i(\tau,x)\label{eomU}
\end{align}
The above repeated indices are not summed up. Subsequently, the conjugate momentum fields $\pi$ and $F_i$ are updated via:
\begin{align}
    \tr[\pi(\tau,x)\sigma^a] &= c_\mathrm{CN}\tr[\pi(\tau-0,x)\sigma^a] + \Delta \tau_\mathrm{eff}\,\bigg\{\tr\bigg[\frac{a^2}{\Delta x^2}\sum_i\left[U_i(\tau,x)\phi(\tau,x+i)U^\dagger_i(\tau,x)\right. \notag\\
    &~~~\left.- 2\phi(\tau,x) + U_i^\dagger(\tau,x-i)\phi(\tau,x-i)U_i(\tau,x-i)\right]\sigma^a\bigg]-a^4\frac{\partial\mathcal{V}}{\partial\phi^a}\bigg\}\;,\label{eompi}
\end{align}
\begin{align}
    \tr\big[\ii\sigma^a F_i(\tau,x)\big] =&\ 
    c_\mathrm{CN}\tr\big[\ii\sigma^aF_i(\tau-0,x)\big] + \Delta \tau_\mathrm{eff}\,\mathrm{Re}\,\bigg\{\frac{a^2}{\Delta x}\,2\tr\left[
            g(\tau)U_i(\tau,x)\phi(\tau,x+i)U_i^\dagger(\tau,x)\ii \sigma^a\phi(\tau,x)\right] \notag\\
       & -\frac{1}{g(\tau)\Delta x^3}
             \tr\sum_j \ii\sigma^a\left[
             U_i(\tau,x)U_j(\tau,x+k)U_i^\dagger(\tau,x+j)U_j^\dagger(\tau,x)\right. \notag\\
            &\ \quad\quad\qquad\quad\qquad~~~
            \left.+ U_i(\tau,x)U_j^\dagger(\tau,x-j+i)U_i^\dagger(\tau,x-j)U_j(\tau,x-j)
            \right]\bigg\}\;,\label{eomF}
\end{align}
where 
\begin{align}
    c_\mathrm{CN} = \frac{2 - \sigma\Delta\tau}{2 + \sigma\Delta\tau}\;,\quad
    \Delta \tau_\mathrm{eff} = \frac{2\Delta \tau}{2+\sigma\Delta\tau}\;,
\end{align}
are the Crank-Nicolson coefficient and effective time interval, respectively. While the phase transition is ongoing, we set the damping coefficient $\sigma = 1.5$ to allow the system to transition smoothly from a bubble configuration to a monopole configuration. Once the phase transition completes, we set $\sigma = 0$, at which point $c_\mathrm{CN} = 1,\ \Delta \tau_\mathrm{eff} = \Delta \tau$. This method ensures that the Gauss constraint, 
\begin{align}
    \mathcal{G}^a(\tau,x)=&\ \frac{1}{\Delta x}\tr
    \sum_i\ii\sigma^a\big[F_i(\tau,x)-U^\dagger_i(\tau,x-i)F_i(\tau,x-i)U_i(\tau,x-i)\big]
    -2\tr\big[
        g\pi(\tau,x)\ii\sigma^a\phi(\tau,x)\big]\;, 
\end{align}
remains well satisfied throughout the simulation.

\section{Identify monopoles on lattice}
After the SU (2) symmetry is broken, under unitary gauge, the diagonal term of the gauge field is exactly the gauge field corresponding to the remaining U(1) symmetry. By using the projection operator~\cite{Hindmarsh_Rummukainen_Weir_2017, Davis_2000, Edwards_2009}
\begin{align}
    P_\pm=\frac12(1_{2\times2}\pm\sigma^3)\;,
\end{align}
we can obtain the U(1) link fields under unitary gauge
\begin{align}
    \tilde V_{i,\pm}(\tau,x)=P_\pm \tilde U_{i}(\tau,x)P_\pm\;.
\end{align}
Only the upper left (lower right) element of $\tilde V_{i,+}\ (\tilde V_{i,-})$ is non-zero. 
The theory \eqref{L} is invariant under gauge transformations
\begin{align}
\phi(\tau,x)&\rightarrow\Lambda(\tau,x)\phi(\tau,x)\Lambda^\dagger(\tau,x)\;,\\
U_i(\tau,x)&\rightarrow\Lambda(\tau,x)U_i(\tau,x)\Lambda^\dagger(\tau,x+i)\;.
\end{align}
So, for any SU(2) triplet under any gauge, the following specification transformation matrix $R(\tau,x)$ can transform $\hat\phi$ at any point into $\sigma^3$.
\begin{align}
    R(\tau,x)\propto\ii[\sigma^3+\hat\phi(\tau,x)]\;,
\end{align}
where $\hat\phi = \sqrt2\phi/\sqrt{\tr(\phi\phi)}$ is the unit SU(2) triplet field.
\begin{align}
    \hat\phi&\rightarrow R\phi R^\dagger\notag\\
    &=(\sigma^3+\hat\phi)\hat\phi(\sigma^3+\hat\phi)\notag\\
    &=\sigma^3\hat\phi\sigma^3+\hat\phi\hat\phi\hat\phi+\sigma^3\hat\phi\hat\phi+\hat\phi\hat\phi\sigma^3\notag\\
    &=(-\hat\phi^1\sigma^1-\hat\phi^2\sigma^2+\hat\phi^3\sigma^3)+(\hat\phi^1\sigma^1+\hat\phi^2\sigma^2+\hat\phi^3\sigma^3) + 2\sigma^3\notag\\
    &=2(1+\hat\phi^3)\sigma^3\notag\\
    &\propto\sigma^3\;,
\end{align}
here we use $\hat\phi\hat\phi=\mathbb{1}$.
So, for SU(2) gauge field $U_i(\tau,x)$ under any gauge, we can obtain the remaining U(1) gauge field $V_i(\tau,x)$ under unitary gauge through $P_\pm$ and $R(\tau,x)$: 
\begin{align}
    \tilde V_{i,\pm}(\tau,x)&=P_\pm R(\tau,x)U_i(\tau,x)R^\dagger(\tau,x+i)P_\pm\notag\\
    &\propto\frac14(\mathbb{1}\pm\sigma^3)[\sigma^3+\hat\phi(\tau,x)]U_i(\tau,x)[\sigma^3+\hat\phi(\tau,x+i)](\mathbb{1}\pm\sigma^3)\notag\\
    &=\frac14[\sigma^3+\hat\phi(\tau,x)\pm\mathbb{1}\pm\sigma^3\hat\phi(\tau,x)]U_i(\tau,x)[\sigma^3+\hat\phi(\tau,x)\pm\mathbb{1}\pm\hat\phi(\tau,x)\sigma^3]\notag\\
    &=[\sigma^3+\hat\phi(\tau,x)]\frac12[\mathbb{1}\pm\hat\phi(\tau,x)]U_i(\tau,x)\frac12[\mathbb{1}\pm\hat\phi(\tau,x+i)][\sigma^3+\hat\phi(\tau,x+i)]\notag\\
    &\propto R(\tau,x)\Pi_\pm(\tau,x)U_i(\tau,x)\Pi_\pm(\tau,x+i)R^\dagger(\tau,x+i)\notag\\
    &= R(\tau,x)V_{i,\pm}(\tau,x)R^\dagger(\tau,x+i)\;.
\end{align}
Now, we have the U(1) link fields under any gauge
\begin{align}
    V_{i,\pm}(\tau,x) = \Pi_\pm(\tau,x)U_i(\tau,x)\Pi_\pm(\tau,x+i)\;,
\end{align}
where the projectors are
\begin{align}
    \Pi_\pm(\tau,x)=\frac12[1_{2\times2}\pm\hat\phi(\tau,x)]\;,\quad \Pi_\pm^2=\Pi_\pm\;.
\end{align}
With the link field, we can define a plaquette
\begin{align}
    V_{ij,\pm}(\tau,x) = V_{i,\pm}(\tau,x)V_{j,\pm}(\tau,x+i)V_{i,\pm}^\dagger(\tau,x+j)V_{j,\pm}^\dagger(\tau,x)\;.\label{plaqV}
\end{align}
Here $V_{ij,+}(\tau,x)$ is still a matrix, but only the upper left element is non-zero. The relationship between its trace and the U(1) gauge field strength $A_{ij}(\tau,x)$ is
\begin{align}
    \tr\, V_{ij,+}(\tau,x) = \exp[\ii g\Delta x^2A_{ij}(\tau,x)/2]\;.
\end{align}
Therefore, the simplest way is to take the argument to obtain the U(1) gauge field strength, which is the magnetic field $B_i$,
\begin{align}
    A_{ij}(\tau,x) &= \frac{2}{g\Delta x^2}\arg\tr\,V_{ij,+}(\tau,x)\;,\\
    B_i(\tau,x) &= \frac12\varepsilon_{ijk}A_{jk}(\tau,x)\;.
\end{align}
Finally, utilizing Gauss's theorem $\nabla\cdot\boldsymbol{B}=\rho_M$, we get the monopole density
\begin{align}
    \frac{1}{\Delta x}\sum_i[B_i(\tau,x+i)-B_i(\tau,x)] = \rho_{M,\bar M}(\tau,x) = \frac{4\pi}{g}n_{M,\bar M}(\tau,x)\;,
\end{align}
where $g/4\pi$ is the magnetic charge of a 't Hooft-Polyakov monopole~\cite{HOOFT1974276}. $n(\tau,x)$ represents the monopole (or antimonopole if it is negative) number density in a lattice site $\Delta x^3$. The total number of monopoles and antimonopoles in a lattice site is
\begin{align}
    N_M(\tau,x) &= n_M(\tau,x)\Delta x^3\;,\quad n_M>0\;,\\
     N_{\bar M}(\tau,x) &=  n_{\bar M}(\tau,x)\Delta x^3\;,\quad n_{\bar M}<0\;,
\end{align}
which are both integers. Sum up the number of monopoles and antimonopoles in each lattice point separately to obtain the total number of monopoles and antimonopoles, and the sum of the two is always $0$ because there are no monopoles or antimonopoles in the lattice at the beginning of the simulation.
\begin{align}
    N_M(\tau) = \sum_x N_M(\tau,x)\;,\quad
    N_{\bar M}(\tau) =  \sum_x N_{\bar M}(\tau,x)\;,\quad
    N_M(\tau) + N_{\bar M}(\tau)\equiv0\;.
\end{align}

\section{The Phase Transition Progress}
\begin{figure}
    \centering
    \includegraphics[width=0.45\linewidth]{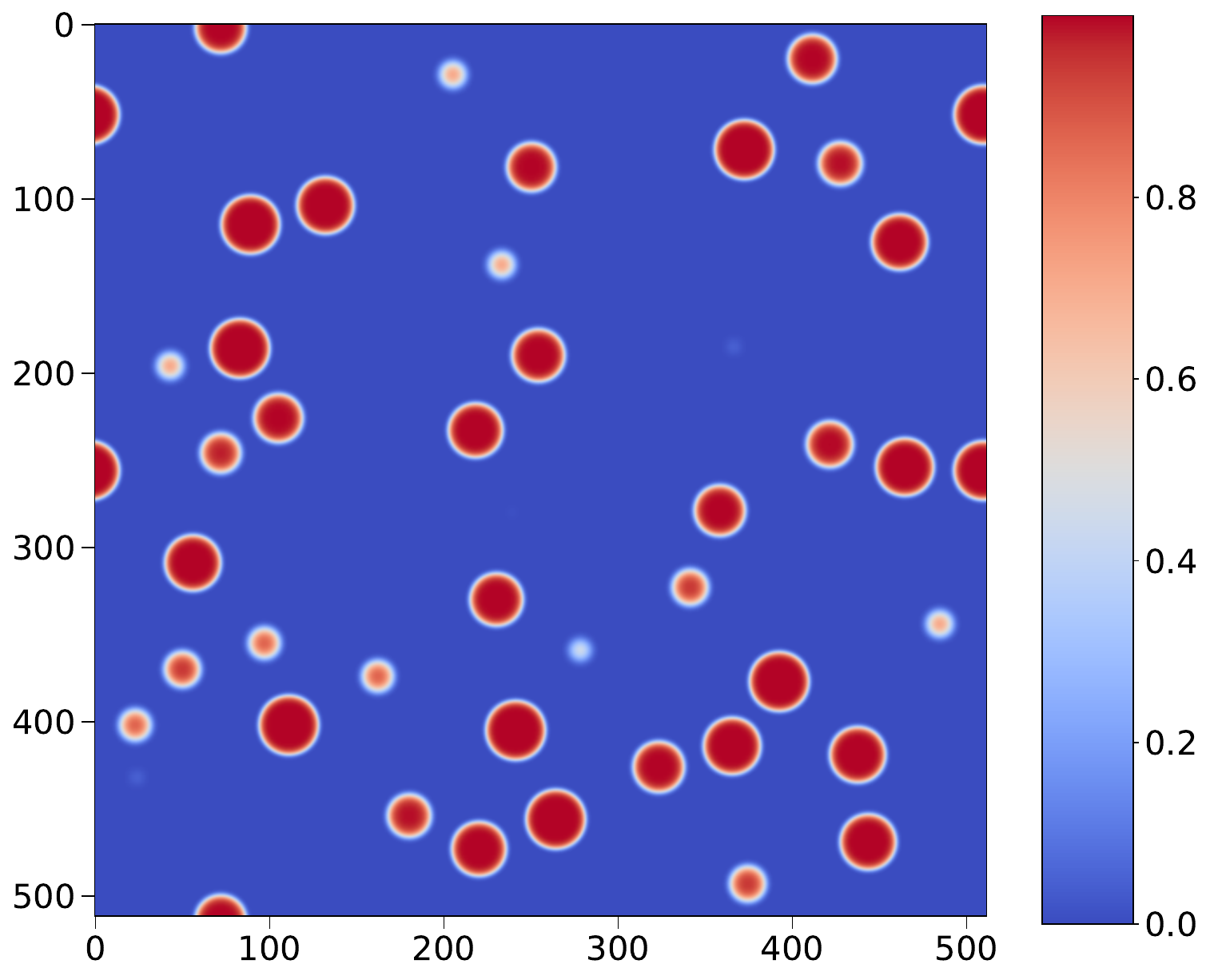}
    \includegraphics[width=0.45\linewidth]{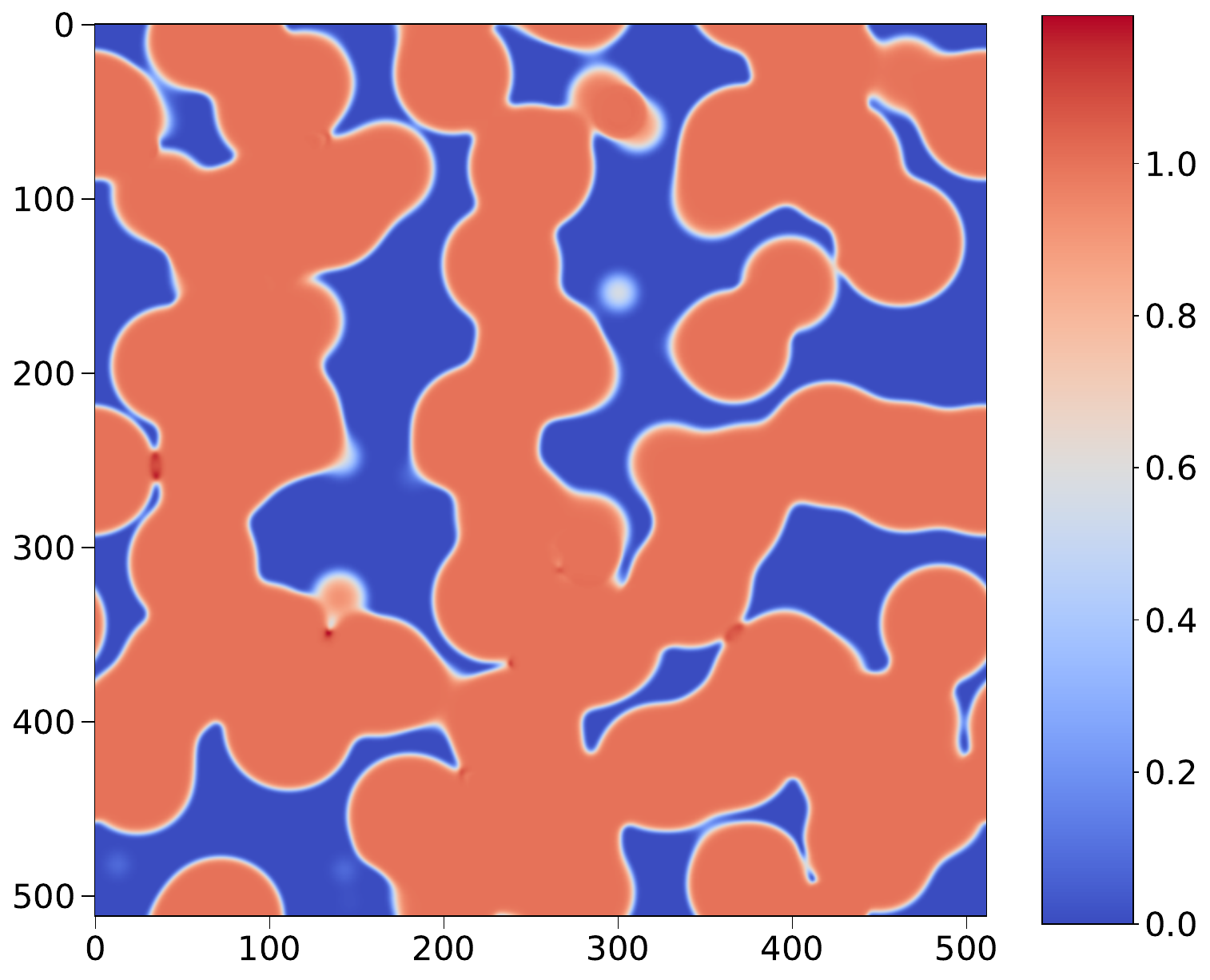}\\
    \includegraphics[width=0.45\linewidth]{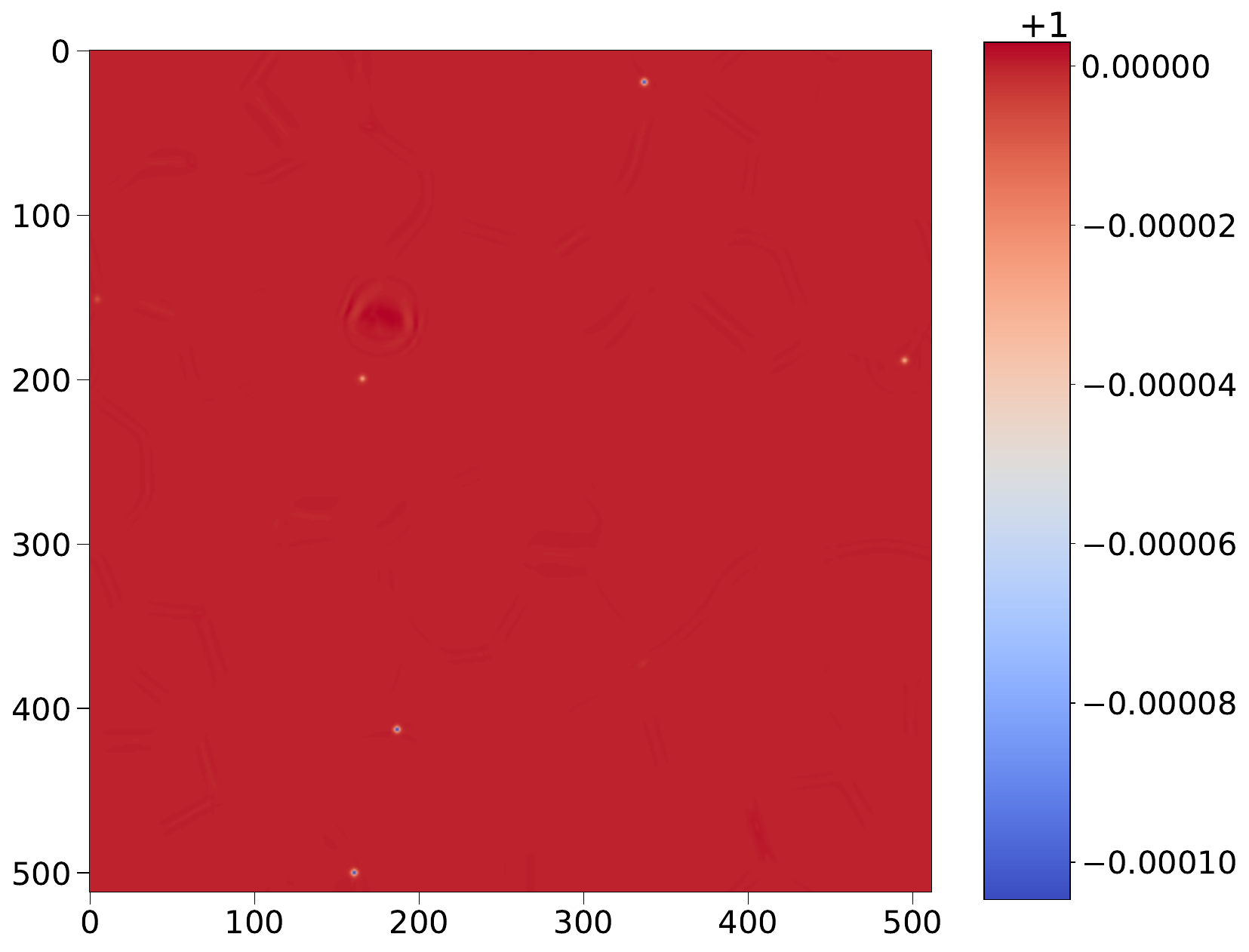}
    \includegraphics[width=0.45\linewidth]{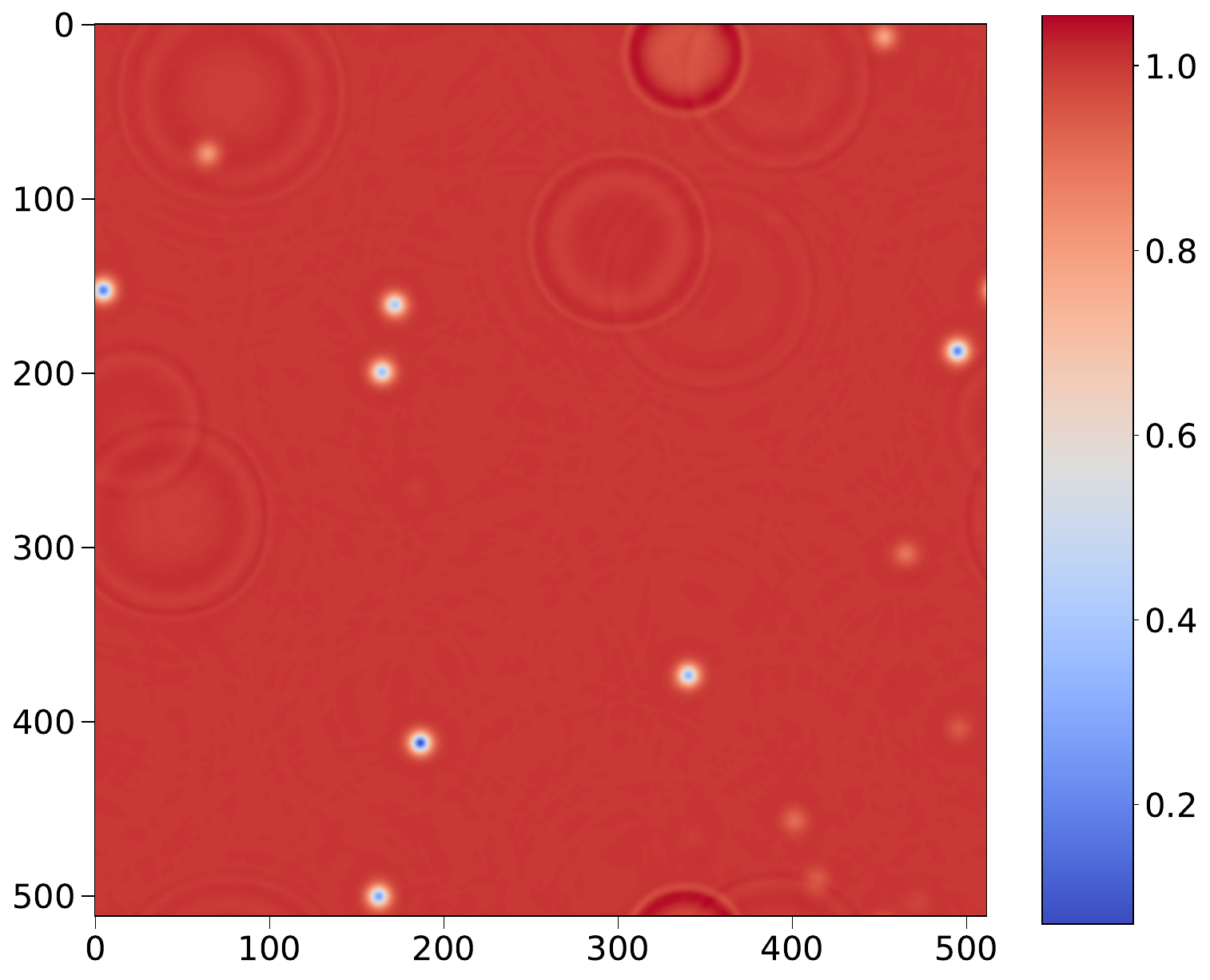}
    \caption{Two-dimensional spatial slices of the scalar field modulus squared $\phi^a\phi^a$ at four representative epochs for a total bubble number $N_b = 500$: bubble nucleation (top left), during the phase transition (top right), at the completion of the phase transition (bottom left), and at the end of the ``core growth" stage (bottom right). The color scale indicates the vacuum state, with blue denoting the false vacuum ($\phi^a\phi^a = 0$) and red representing the true vacuum ($\phi^a\phi^a \approx v$). The bottom left panel presents the deviation of $\phi^a\phi^a$ from ``+1". In the bottom right panel, the cores of the formed monopoles remain trapped in the false vacuum state ($\phi^a\phi^a = 0$), while the surrounding background has fully relaxed into the true vacuum.}
    \label{fig:slice}
\end{figure}
In this Appendix, we briefly demonstrate the dynamics of the FOPT. At the onset of the simulation, we simultaneously nucleate $N_b$ bubbles within the false-vacuum $\phi^a\phi^a=0$ background, with the initial mean bubble separation given by $R_*=(V/N_b)^{1/3}$, as depicted in the top-left panel of Fig. \ref{fig:slice}. Evolving under the equations of motion (\ref{eomphi}$\sim$\ref{eomF}), the bubbles expand, collide, and merge with one another, as illustrated in the top-right panel of Fig. \ref{fig:slice}. Once the entire volume relaxes into the true vacuum $\phi^a\phi^a\approx v$, the phase transition terminates, as shown in the bottom-left panel of Fig. \ref{fig:slice}. Throughout the phase transition, the expansion of the universe is neglected. Concurrently, due to our choice of relatively large initial couplings $\lambda_0$ and $g_0$, the physical size of the monopoles at this stage is smaller than a single lattice spacing $\Delta x$, causing the scalar field $\phi$ at the monopole center to remain close to the true vacuum. Following the completion of the phase transition, the ``core growth" stage of the PRS method commences. In this stage, $\lambda$ and $g$ decrease as the scale factor $a$ increases, allowing the monopole cores to broaden and span across multiple lattice sites. The bottom-right panel displays the distribution of $\phi^a\phi^a$ at the conclusion of the ``core growth" stage, where the false vacuum remains trapped at the monopole cores while the rest of space is occupied by the true vacuum.

\begin{figure}
    \centering
    \includegraphics[width=0.45\linewidth]{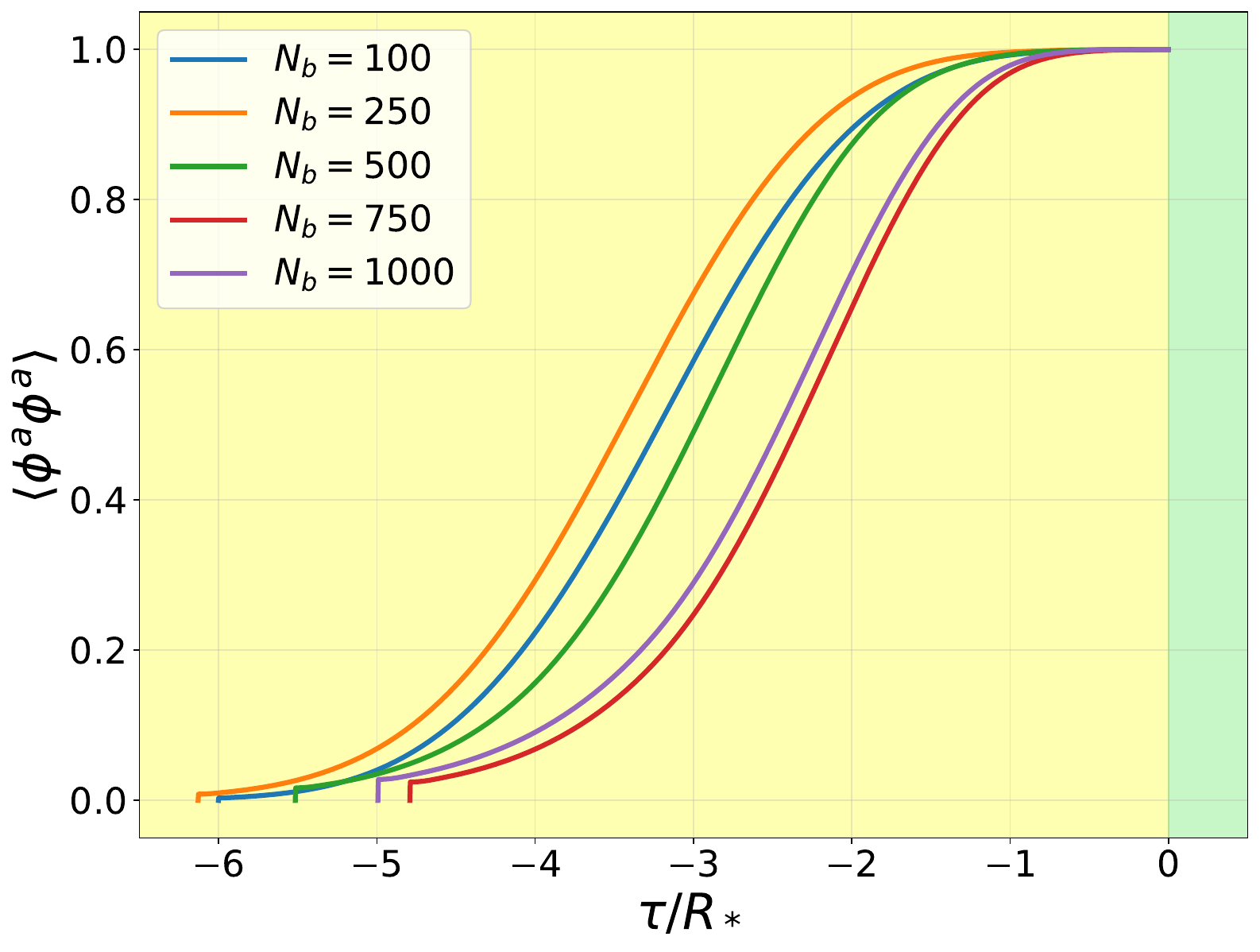}
    \includegraphics[width=0.45\linewidth]{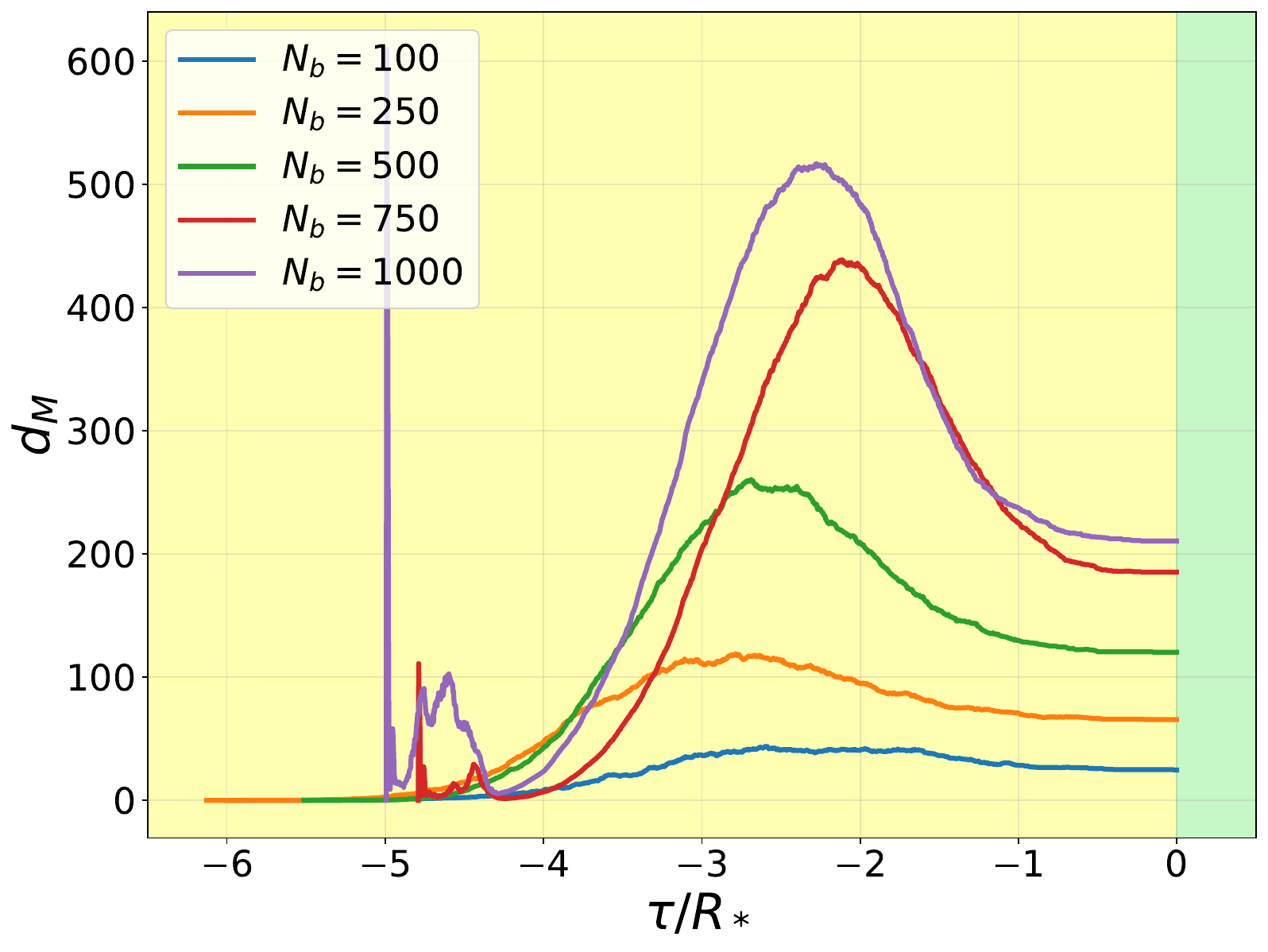}
    \caption{Left: Time evolution of the squared modulus of the scalar field $\phi$ for various bubble numbers $N_b$, characterizing the progression of the phase transition. Right: Time evolution of the monopole number $N_M$ during the phase transition for different $N_b$. On the horizontal axis, $R_*=(V/N_b)^{1/3}$ denotes the mean bubble separation. The expansion begins at $\tau = 0$, so the phase transition proceeds entirely in the regime $\tau < 0$.}
    \label{fig:phi2}
\end{figure}
Fig. \ref{fig:phi2} illustrates the temporal evolution of $\phi^a\phi^a$ (Left panel) and the monopole number $N_M$ (Right panel) throughout the phase transition. It should be emphasized again that the onset of the expansion is defined as $\tau = 0$; consequently, the phase transition proceeds entirely in the regime $\tau < 0$. Because of damping $\sigma=1.5$, the progression of the phase transition is slowed down; typically, the transition completes around $\tau \sim R_*$. Following the nucleation of monopoles via bubble collisions, closely separated monopole–antimonopole pairs rapidly annihilate, leading to an initial reduction in $N_M$. However, because we adopt relatively large initial couplings $\lambda$ and $g$, the monopole core size remains smaller than a single lattice spacing $\Delta x$. This severely restricts the free motion of the monopoles (lattice pinning), causing $N_M$ to plateau near the end of the phase transition stage. It is not until the ``core growth" stage broadens the monopole cores across multiple lattice sites that the monopoles regain mobility and resume annihilation.

\section{The Time Evolution of the Power Spectrum}
\begin{figure*}[!htp]
    \centering
    \includegraphics[width=0.45\linewidth]{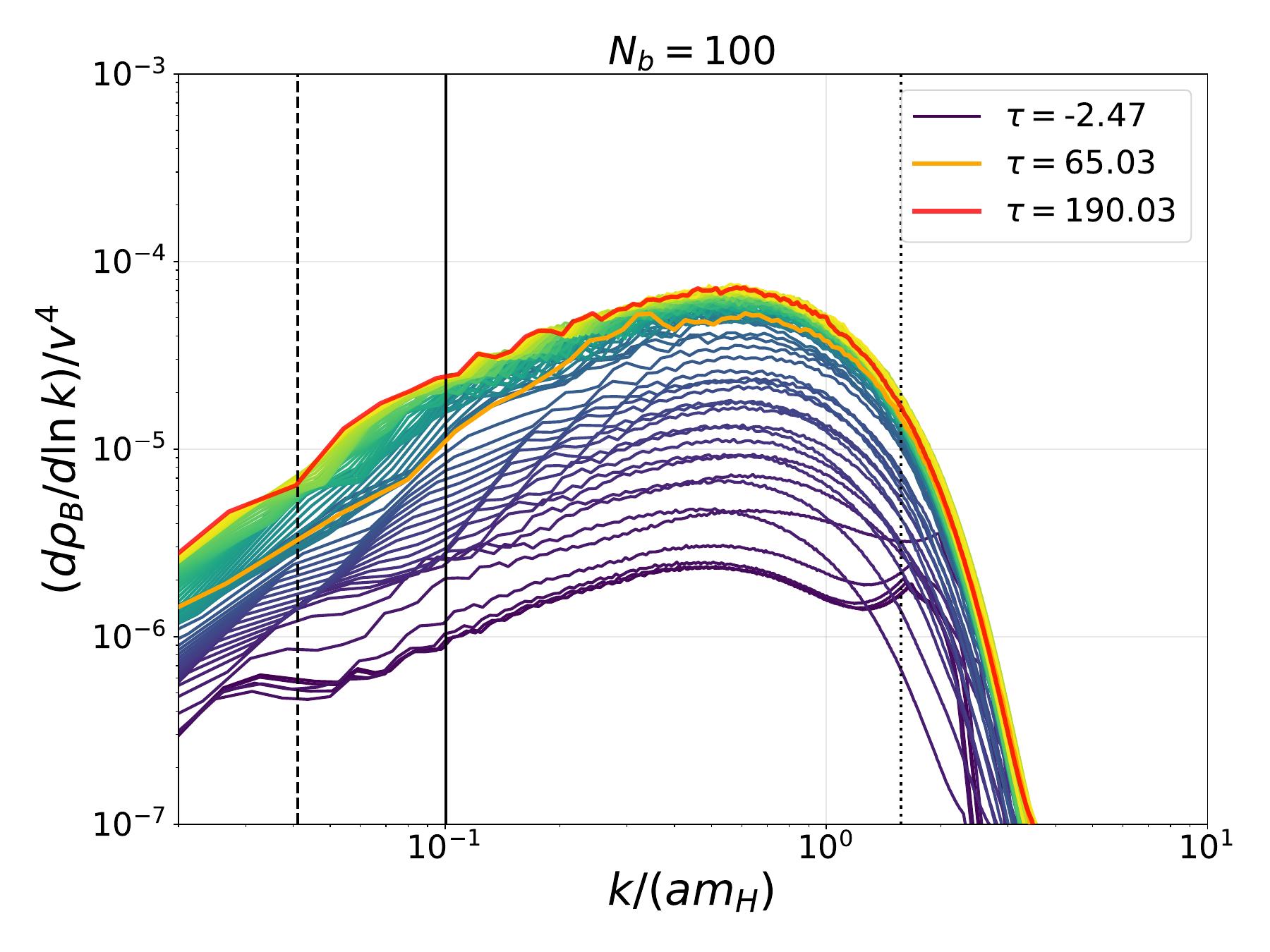}
    \includegraphics[width=0.45\linewidth]{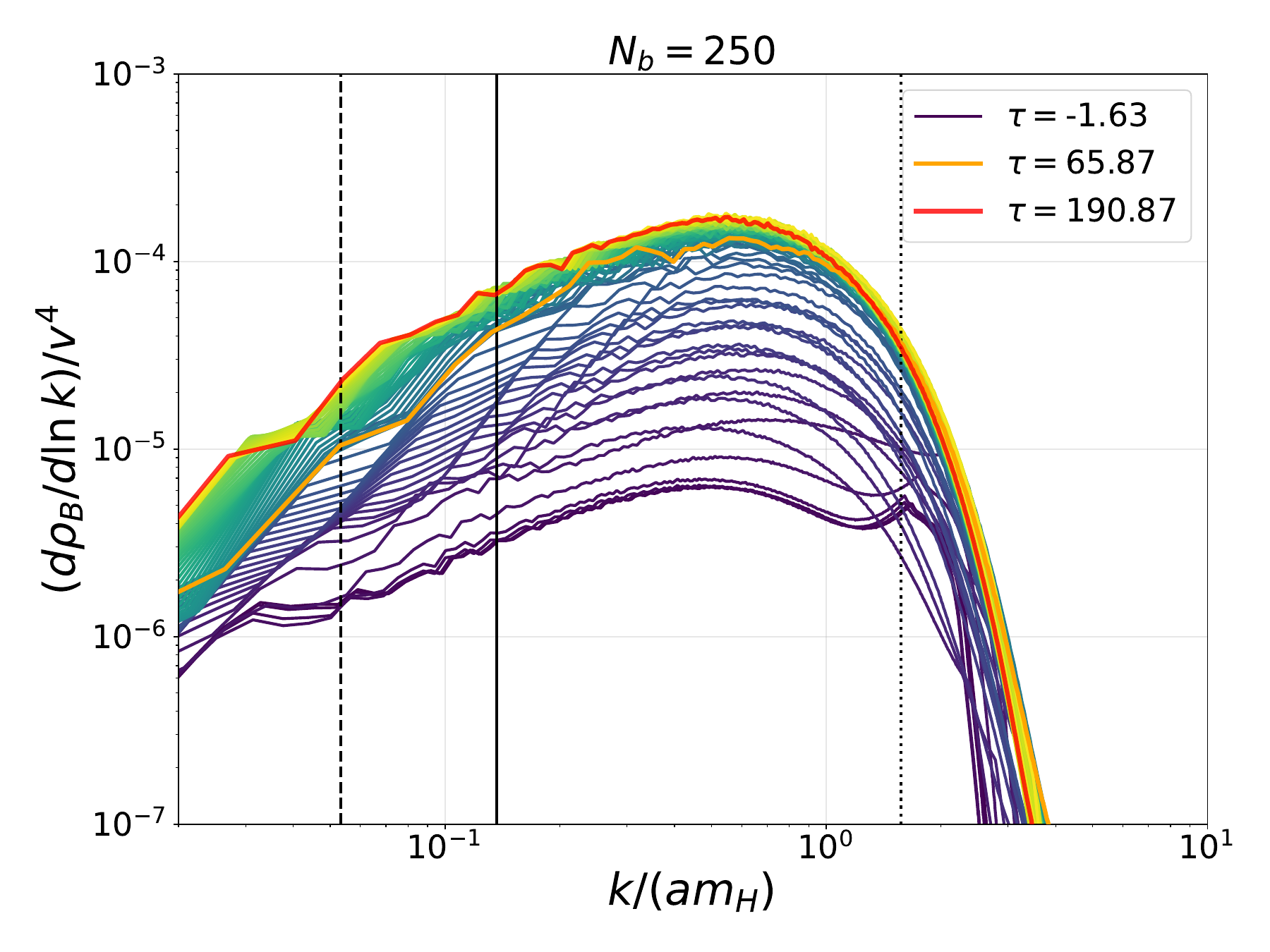}\\
    \includegraphics[width=0.45\linewidth]{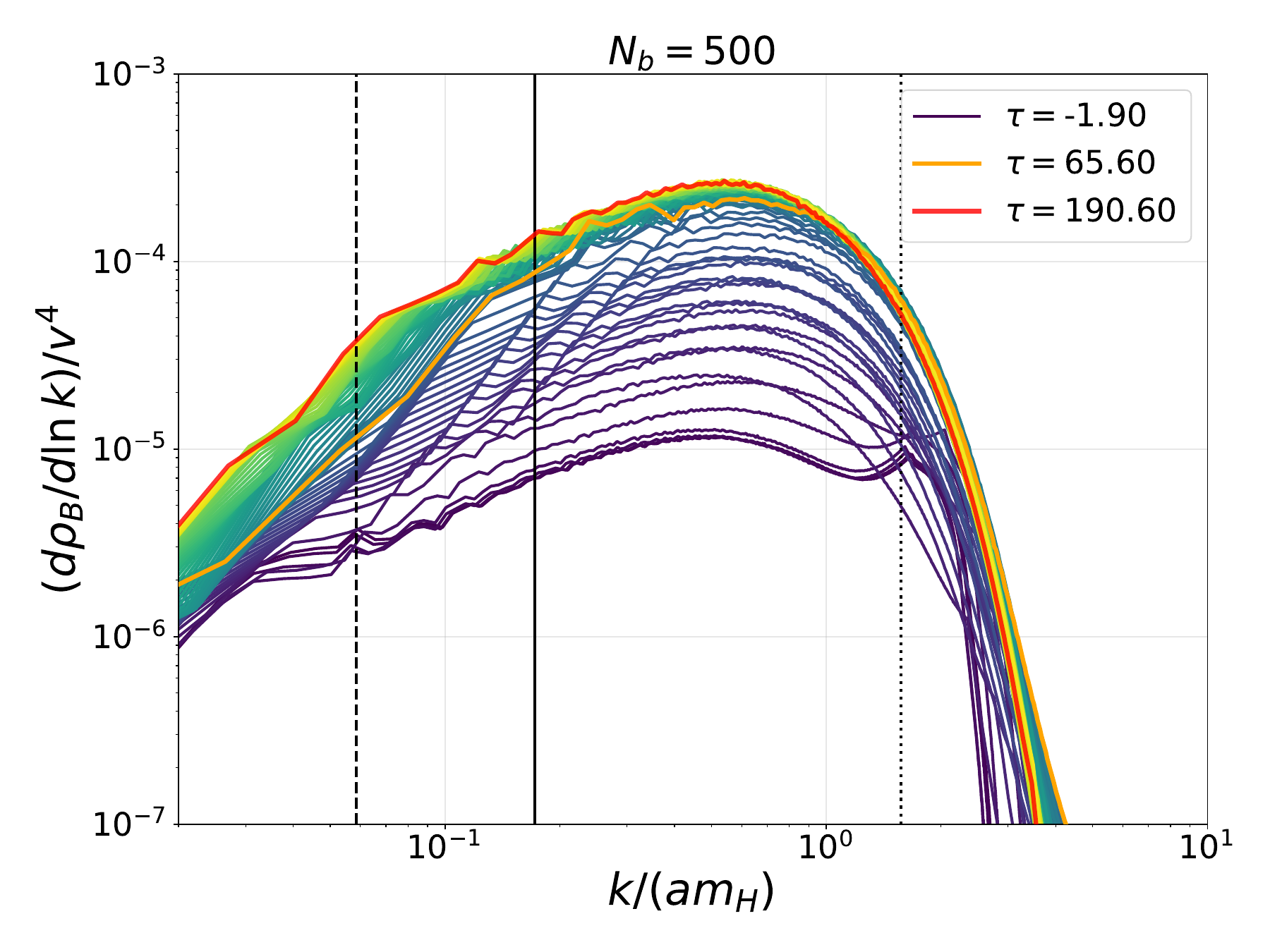}
    \includegraphics[width=0.45\linewidth]{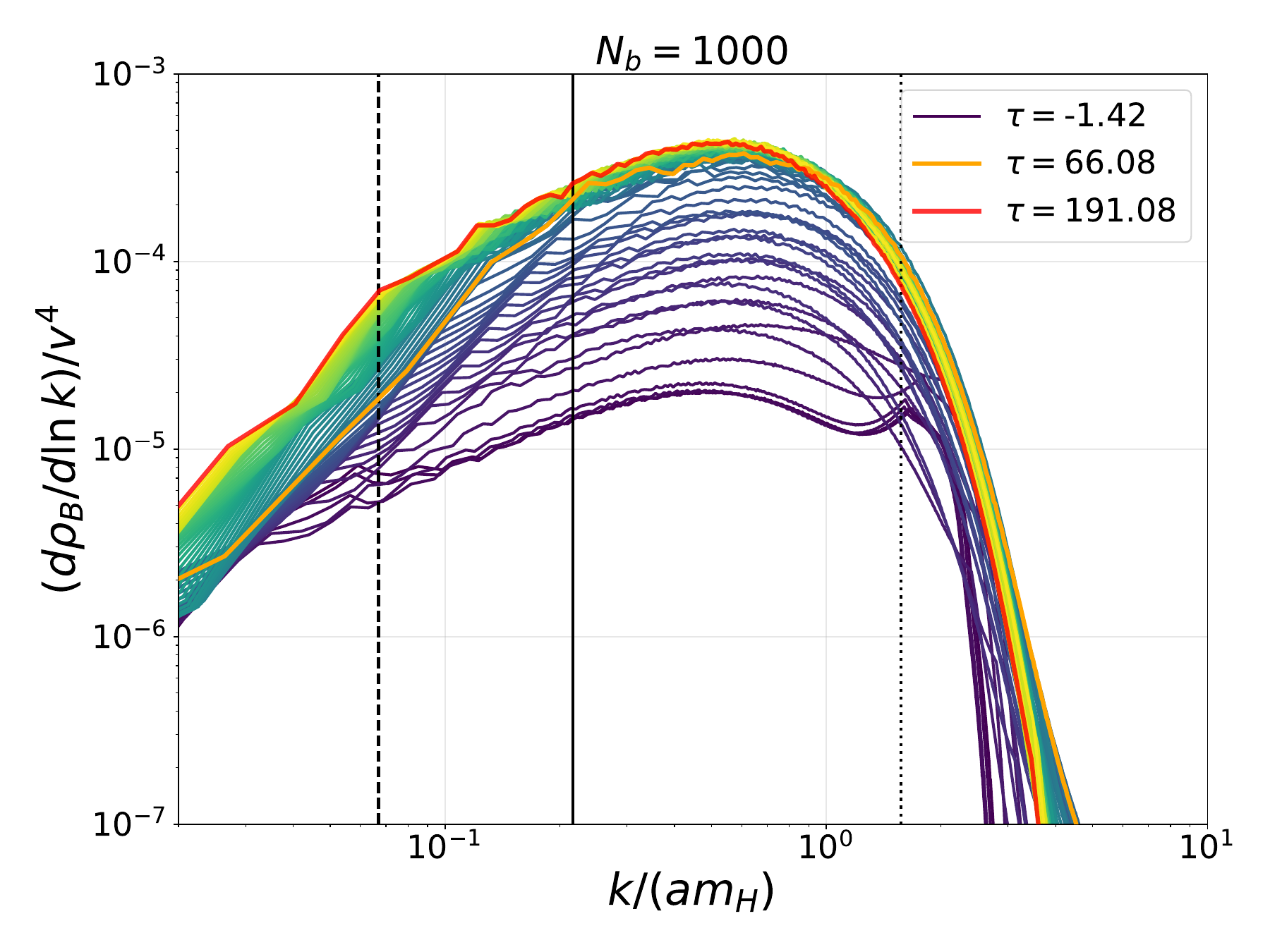}
    \caption{Evolution of the magnetic field spectrum $\dd\rho_B/\dd\ln k$ for different bubble numbers $N_{b}$. The color gradient from purple to yellow denotes the physical chronological evolution from earlier to later times. The orange and red solid lines highlight the spectrum at the end of the ``core growth" stage and the whole simulation, respectively. The black solid, dashed, and dotted vertical lines mark the physical characteristic scales $k/a = 2\pi/R_*$, $k/a = 2\pi/d_M$, and $k/a = 2\pi/w_M$ at the end of the simulation, respectively. }
    \label{fig:B2}
\end{figure*}

Fig.~\ref{fig:B2} shows the time evolution of the magnetic field after the phase transition for various values of $N_b$. Aside from the evident gradual amplification of the magnetic field spectra, their peak positions shift left and right during expansion. This behavior originates from the distinct dependencies of the monopole core size $w_M=1/m_H$ on the scale factor $a$ during the 'core growth' stage versus the subsequent physical evolution stage. This effect will be seen even more clearly in the gravitational wave power spectra below. 

\begin{figure*}
    \centering
    \includegraphics[width=0.45\linewidth]{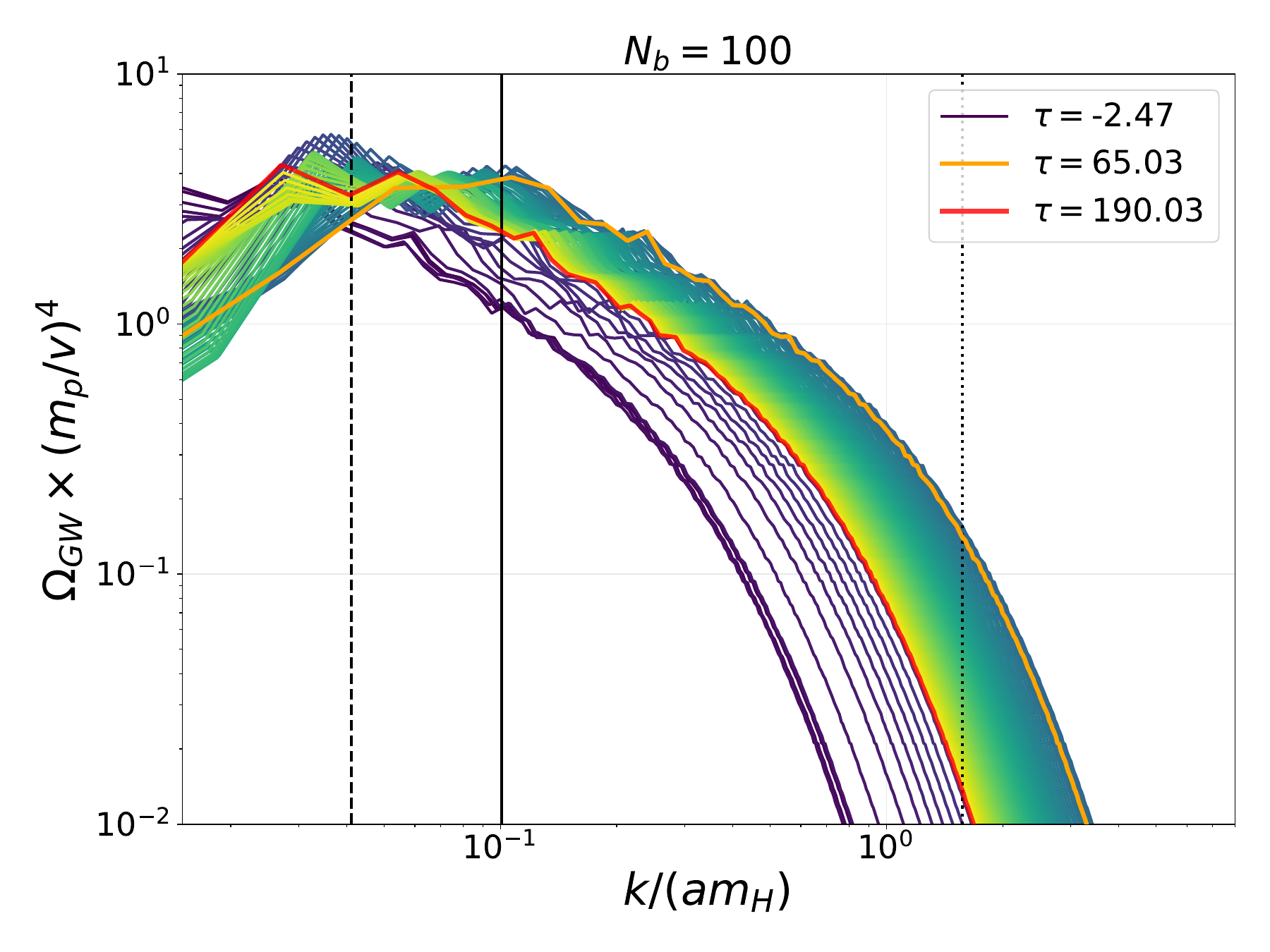}
    \includegraphics[width=0.45\linewidth]{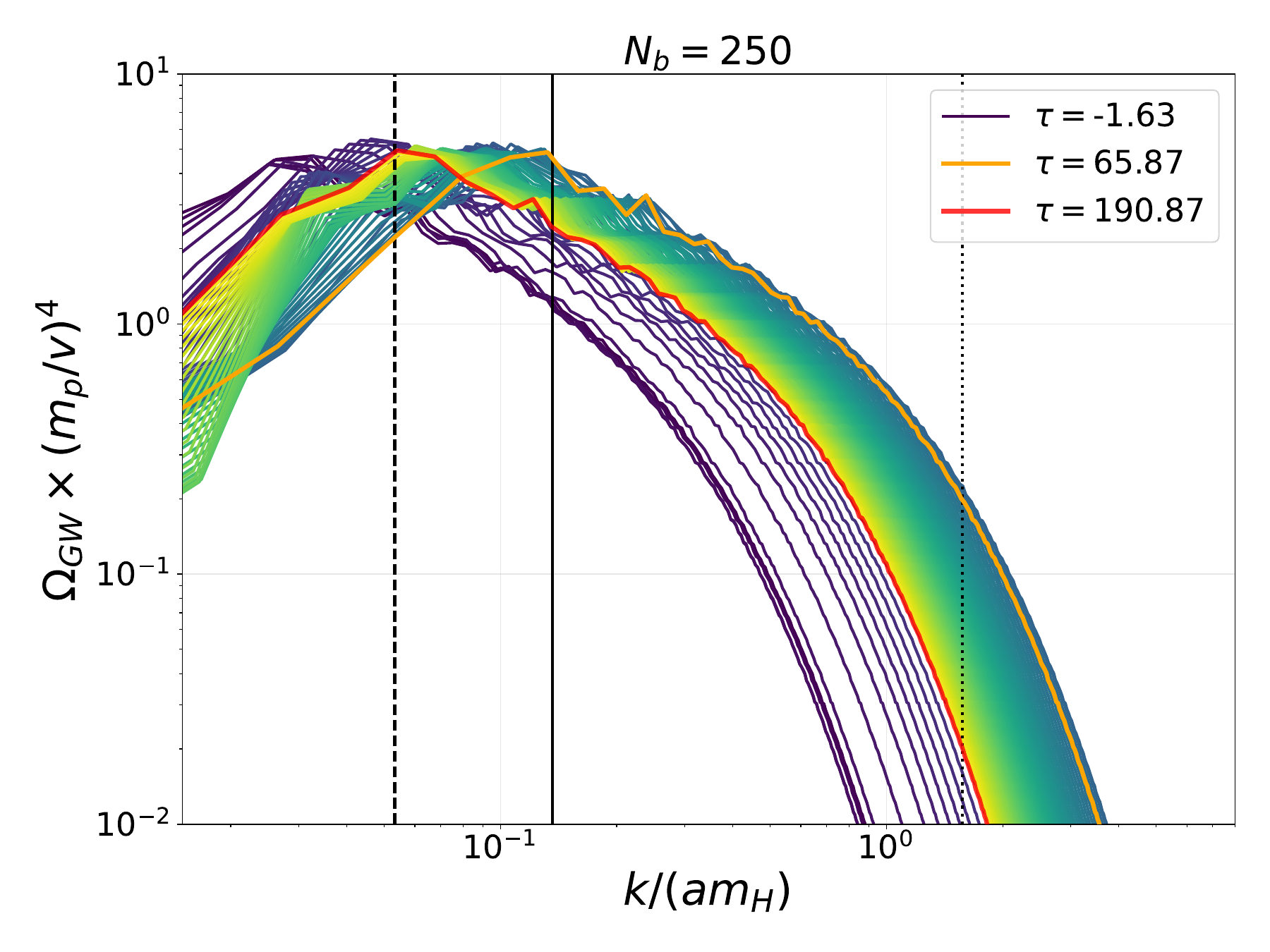}\\
    \includegraphics[width=0.45\linewidth]{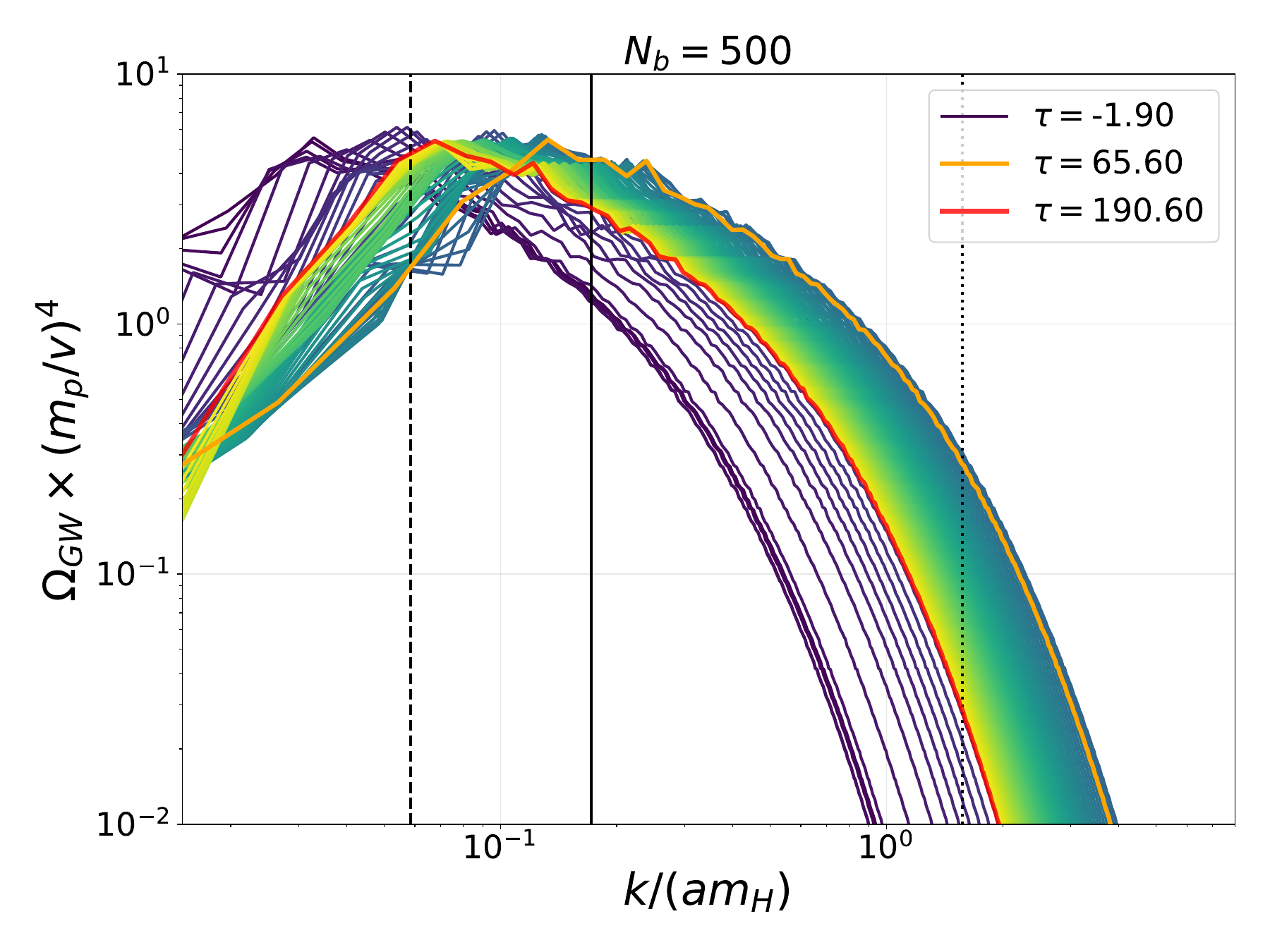}
    \includegraphics[width=0.45\linewidth]{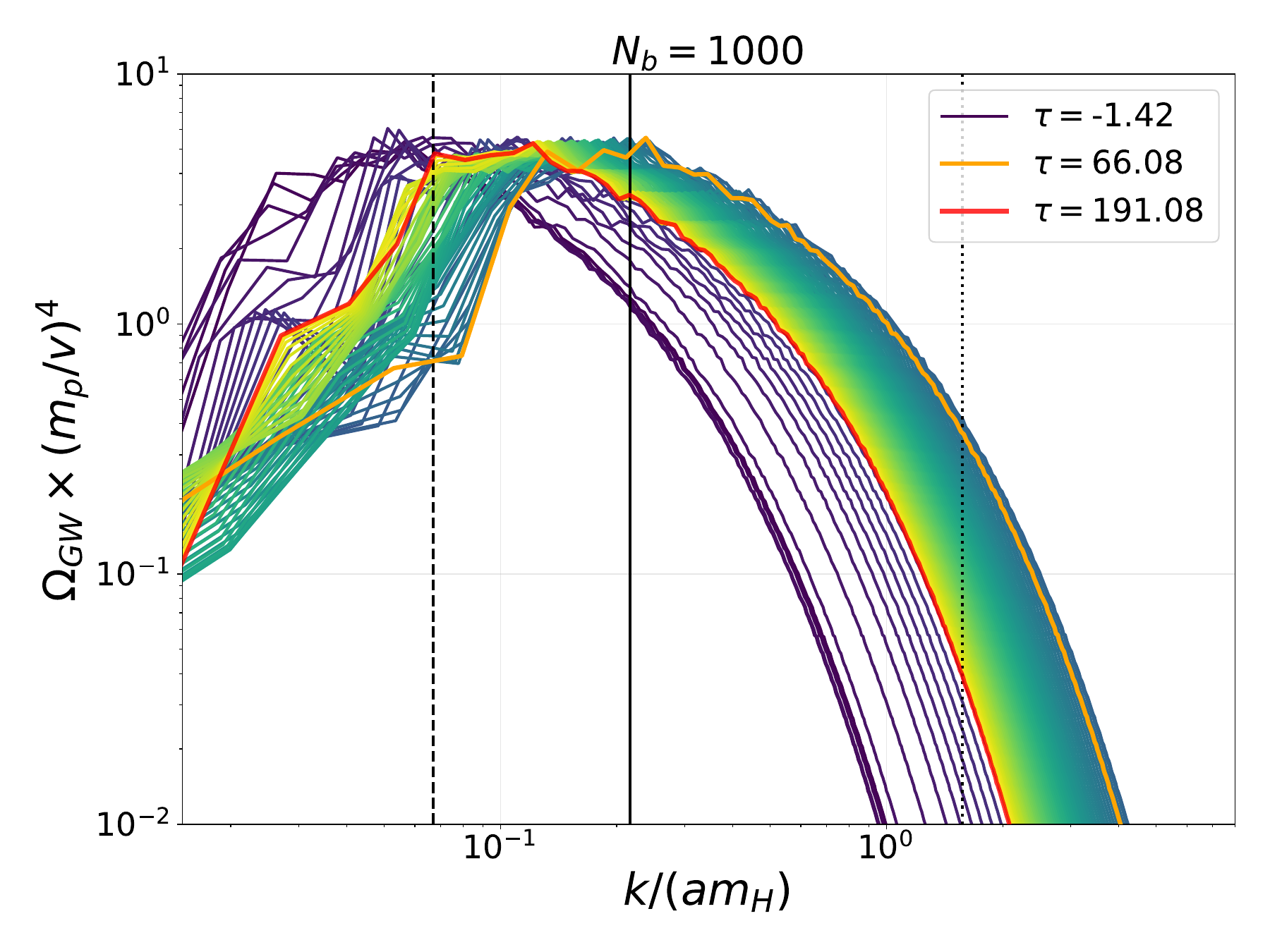}
    \caption{Evolution of the gravitational wave power spectrum $\Omega_\mathrm{GW}$ for different bubble numbers $N_{b}$. The color gradient from purple to yellow denotes the chronological evolution from earlier to later times. The orange and red solid lines highlight the spectrum at the end of the ``core growth" stage and the whole simulation, respectively. The black solid, dashed, and dotted vertical lines mark the physical characteristic scales $k/a = 2\pi/R_*$, $k/a = 2\pi/d_M$, and $k/a = 2\pi/w_M$ at the end of the simulation, respectively. }
    \label{fig:OmegaGW}
\end{figure*}
Fig.~\ref{fig:OmegaGW} shows the time evolution of the magnetic field after the phase transition for various values of $N_b$. The effects of the PRS ``core growth" stage and the expansion are considerably more pronounced in this figure than in the magnetic field spectrum of Fig. \ref{fig:B2}. During the ``core growth" stage, $k/(am_H)\propto a^2$, causing the power spectrum curve to shift from left to right (from the purple curve to the orange curve). Once the ``core growth" stage concludes, $m_H$ ceases to vary while the universe continues to expand, giving $k/(am_H) \propto a^{-1}$; consequently, the power spectrum curve shifts from right to left (from the orange curve to the red curve).

\end{document}